\documentclass{aa}  

\usepackage{graphicx}
\usepackage{txfonts}
\usepackage{caption}
\usepackage{color}
\usepackage{natbib}
\bibpunct{(}{)}{;}{a}{}{,} 
\pdfoutput=1

\begin{document} 

   \title{Probing the use of spectroscopy to determine the meteoritic analogues of meteors}
   \author{A. Drouard
          \inst{1,3}\fnmsep\thanks{\email{alexis.drouard@lam.fr}}
          ,
          P. Vernazza\inst{1}
          ,
          S. Loehle\inst{2}
          ,
          J. Gattacceca\inst{3}
          ,
          J. Vaubaillon \inst{4}
         ,
          B. Zanda \inst{4,5}
          ,
          M. Birlan \inst{4,7}
          ,
          S. Bouley \inst{4,6}
          ,
          F. Colas \inst{4}
          ,
          M. Eberhart \inst{2}
          ,
          T. Hermann\inst{2}
          ,
          L. Jorda \inst{1}          
          ,          
          C. Marmo \inst{6}
          ,
          A. Meindl\inst{2}
          ,
          R. Oefele\inst{2}
          ,
          F. Zamkotsian \inst{1}      
	  \and             
          F. Zander\inst{2}
            }
          \authorrunning{A. Drouard et al.}
          \titlerunning{Probing the use of spectroscopy to determine the meteoritic analogues of meteors}

   \institute{Aix Marseille Université, CNRS, LAM (Laboratoire d'Astrophysique de Marseille) UMR 7326, F-13388, Marseille, France
         \and
            IRS, Universität Stuttgart, Pfaffenwaldring 29, D–70569 Stuttgart, Germany
         \and
             Aix-Marseille Université, CNRS, IRD, Coll France, CEREGE  UM34, F-13545 Aix en Provence, France
         \and
            IMCCE, Observatoire de Paris, Paris, France
         \and 
            Université Pierre et Marie Curie Paris, IMPMC-MNHN, Paris, France
         \and 
           GEOPS, Univ. Paris-Sud, CNRS, Université Paris-Saclay, Rue du Belvédère, Bât. 509, 91405, Orsay, France
         \and 
           Astronomical Institute of Romanian Academy, 5, Cutitul de Argint Street, 040557 Bucharest, Romania
             }
   \date{Received November 2, 2017; accepted January 26, 2018}

 
  \abstract
   {Determining the source regions of meteorites is one of the major goals of current research in planetary science. Whereas asteroid observations are currently unable to pinpoint the source regions of most meteorite classes, observations of meteors with camera networks and the subsequent recovery of the meteorite may help make progress on this question. The main caveat of such an approach, however, is that the recovery rate of meteorite falls is low (<20 \%), implying that the meteoritic analogues of at least 80\% of the observed falls remain unknown.}
   {Spectroscopic observations of incoming bolides may have the potential to mitigate this problem by classifying the incoming meteoritic material.}
   {To probe the use of spectroscopy to determine the meteoritic analogues of incoming bolides, we collected emission spectra in the visible range (320-880 nm) of five meteorite types (H, L, LL, CM, and eucrite) acquired in atmospheric entry-like conditions in a plasma wind tunnel at the Institute of Space Systems at the University of Stuttgart (Germany). A detailed spectral analysis including a systematic line identification and mass ratio determinations (Mg/Fe, Na/Fe) was subsequently performed on all spectra.}
   {It appears that spectroscopy, via a simple line identification, allows us to distinguish the main meteorite classes (chondrites, achondrites and irons) but it does not have the potential to distinguish for example an H chondrite from a CM chondrite.}
   {The source location within the main belt of the different meteorite classes (H, L, LL, CM, CI, etc.) should continue to be investigated via fireball observation networks. Spectroscopy of incoming bolides only marginally helps precisely classify the incoming material (iron meteorites only). To reach a statistically significant sample of recovered meteorites along with accurate orbits (>100) within a reasonable time frame (10-20 years), the optimal solution may be the spatial extension of existing fireball observation networks.}

               
   \maketitle
%

\section{Introduction}

Meteorites are a major source of material to contribute to the understanding of how the solar system formed and evolved \citep{Hutchison01}. They are rocks of extraterrestrial origin, mostly fragments of small planetary bodies such as asteroids or comets, some of which may have orbits that cross that of the Earth. However, both dynamical studies and observation campaigns imply that most meteorites have their source bodies in the main asteroid belt and not among the near-Earth asteroids or comets \citep{Vernazza08}. Yet, it is very difficult to conclusively identify the parent bodies of most meteorite groups using ground-based observations and/or spacecraft data alone. This stems from the fact that most asteroids are not spectrally unique. Therefore, there are plenty of plausible parent bodies for a given meteorite class \citep{Vernazza14, Vernazza16}; the obvious exception is (4)\,Vesta, which appears to be the parent body of the Howardite–Eucrite–Diogenite (HED) group \citep{Binzel93, McSween13}.\\

\noindent
One of the approaches to make progress on the fundamental question of where meteorites come from is to determine both the orbit and composition of a statistically significant sample (>100) of meteoroids. This is typically achieved by witnessing their bright atmospheric entry via dense (60-100 $\mathrm{km}$ spacing) camera/radio networks. These networks allow scientists to accurately measure their trajectory from which both their pre-atmospheric orbit (thus their parent body within the solar system) and the fall location of the associated meteorite (with an accuracy of the order of a few kilometres) can be constrained.\\

\noindent
Several camera networks already exist (or have existed) around the world (USA, Canada, Central Europe, and Australia). These networks have allowed researchers to constrain the orbital properties of $\sim$20 meteoroids, which were recovered as meteorites (see \cite{Ceplecha60}, \cite{McCrosky71}, \cite{Halliday81},  \cite{Brown94}, \cite{Brown00}, \cite{Spurny03}, \cite{Borovicka03}, \cite{Trigo04b}, \cite{Simon04}, \cite{Hildebrand09}, \cite{Haack10}, \cite{Jenniskens10}, \cite{Spurn10}, \cite{Brown11}, \cite{Jenniskens12}, \cite{Spurny12}, \cite{Spurny12b}, \cite{Borovicka13}, \cite{Borovicka13b}, \cite{Dyl16}). The main limitation of these networks is their size. Most of these consist of a fairly small number of cameras spread over a comparatively small territory. This implies that the number of bright events per year witnessed by these networks is small and that tens of years would be necessary to properly constrain the source regions of meteorites and yield a significant number (>100) of samples. To overcome this limitation, larger networks have been designed and deployed: FRIPON over France \citep{Colas15}, PRISMA over Italy \citep{Gardiol16} and the Desert Fireball Network over Australia \citep{Bland14}. These networks will ultimately allow us to reduce the number of years necessary to achieve a statistical sample of recovered meteorites with well-characterized orbits. However, an important limit of this approach is the discrepancy between the number of accurate orbits and the number of recovered meteoritic samples. A recovery rate of  $\sim$20\% is perhaps an upper limit implying that the meteoritic analogues for at least 80\% of the bolides observed by the various networks will remain unknown.\\

\noindent
Spectroscopic observations of incoming bolides may have the potential to mitigate this problem by classifying the incoming meteoritic material. However, whereas meteor spectra have been routinely collected in the visible wavelength range for more than three decades \citep{Borovicka93, Borovicka94, Trigo03, Trigo04, Trigo04c, Borovicka05, Borovicka05b, Madiedo14, Kasuga04, Babadzhanov04, Madiedo14b, Vojacek15, Mozgova15, Koukal16, Bloxam17}, and while a few emission spectra of meteorites have been collected experimentally using various techniques (e.g. electrostatic accelerator \citep{Friichtenicht68} or laser \citep{Milley07}), only a few bolide-meteorite associations were proposed - yet not firmly established - based on this technique \citep{Borovika94, Madiedo2013}. In a nutshell, it is still unclear whether spectroscopy can inform us about the nature of incoming bolides.\\
 
\noindent
To probe the use of spectroscopy to determine the meteoritic analogues of incoming bolides, we collected emission spectra of five meteorite types (H, L, LL, CM, and eucrite) acquired in atmospheric entry-like conditions in the plasma wind tunnel at the Institute of Space Systems (IRS) at the University of Stuttgart (Germany). Thus far, a plasma wind tunnel is the only tool that allows us to reproduce atmospheric entry-like heating conditions. Wind tunnels have been used in the past to study the ablation process of meteoroids but emission spectra of the ablated materials were not the main interest of the study \citep{Shepard67}. The first experiment of such kind dedicated to spectroscopy was recently performed on an H chondrite, a terrestrial argillite and a terrestrial basalt \citep{Loehle17}.\\

\noindent
Here, we use our new data along with those acquired by \cite{Loehle17} to determine how much compositional information can be retrieved for a vapourized matter based on its emission spectrum.


\section{Laboratory experiments using a wind tunnel}

\subsection{Methods}

We simulated the atmospheric entry of meteoroids using the plasma wind tunnel PWK1 with the aim of collecting the emission spectra in the visible wavelength range of five meteorite types (eucrite, H, L, LL, and CM; see section 2.2 for more details). The facility at IRS provides an entry simulation of flight scenarios with enthalpies as expected during flight in the upper atmosphere (~70 $\mathrm{MJ\, kg^{-1}}$) and the suite of diagnostic methods (high-speed imaging, optical emission spectroscopy, thermography) allows a comprehensive experimental analysis of the entry simulation in the plasma wind tunnel. The experiments took place during the first week of May 2017.\\

\begin{figure} [h!]
   \centering
   \includegraphics[scale=0.3]{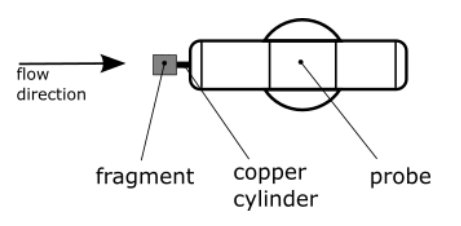}
      \caption{Schematic view (from top) of the sample mounted on the probe.
              }
         \label{probe}
   \end{figure}

\noindent
The experimental procedure was as follows. First, the samples were cut into cylinders (10 $\mathrm{mm}$ diameter, 10 $\mathrm{mm}$ high) and a smaller cylindrical hole (diameter 3 $\mathrm{mm}$, 5 $\mathrm{mm}$ deep) was drilled on one side in order to fix the sample on a copper cylinder fitting the probe (see Fig. \ref{probe}). These rock cutting and drilling steps were undertaken at CEREGE (France) where adequate rock saws (wire saw and low-speed saw) and drills are maintained. Second, the sample and the copper cylinder were fixed onto the probe. The latter is aligned in the flow direction, facing the plasma generator (see Fig. \ref{probe}), and is mounted on a movable platform inside a large vacuum chamber. The size of the chamber prevents flow-wall interactions. 

\begin{figure} [h!]
   \centering
   \includegraphics[width=\hsize]{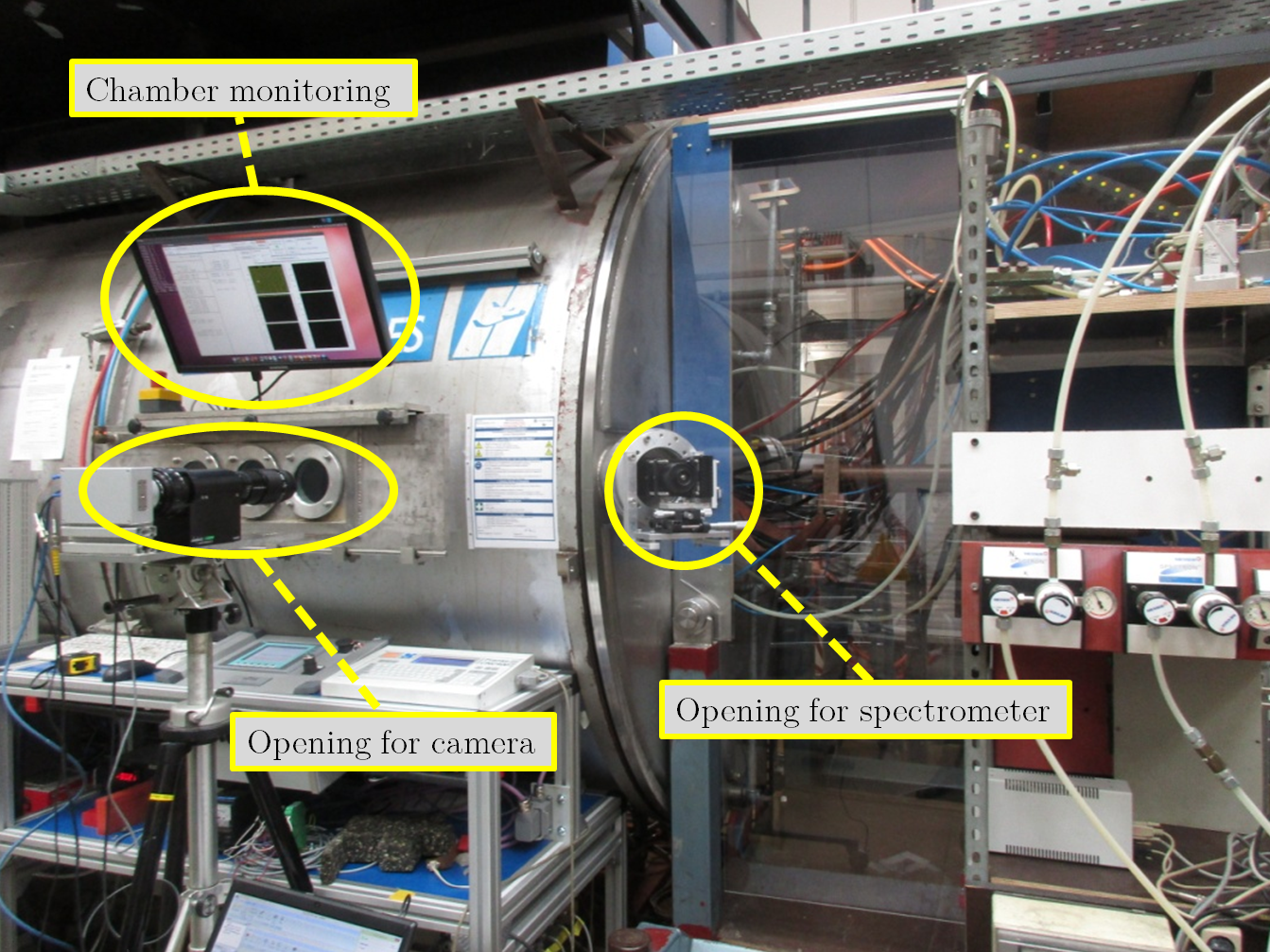}
   \includegraphics[width=\hsize]{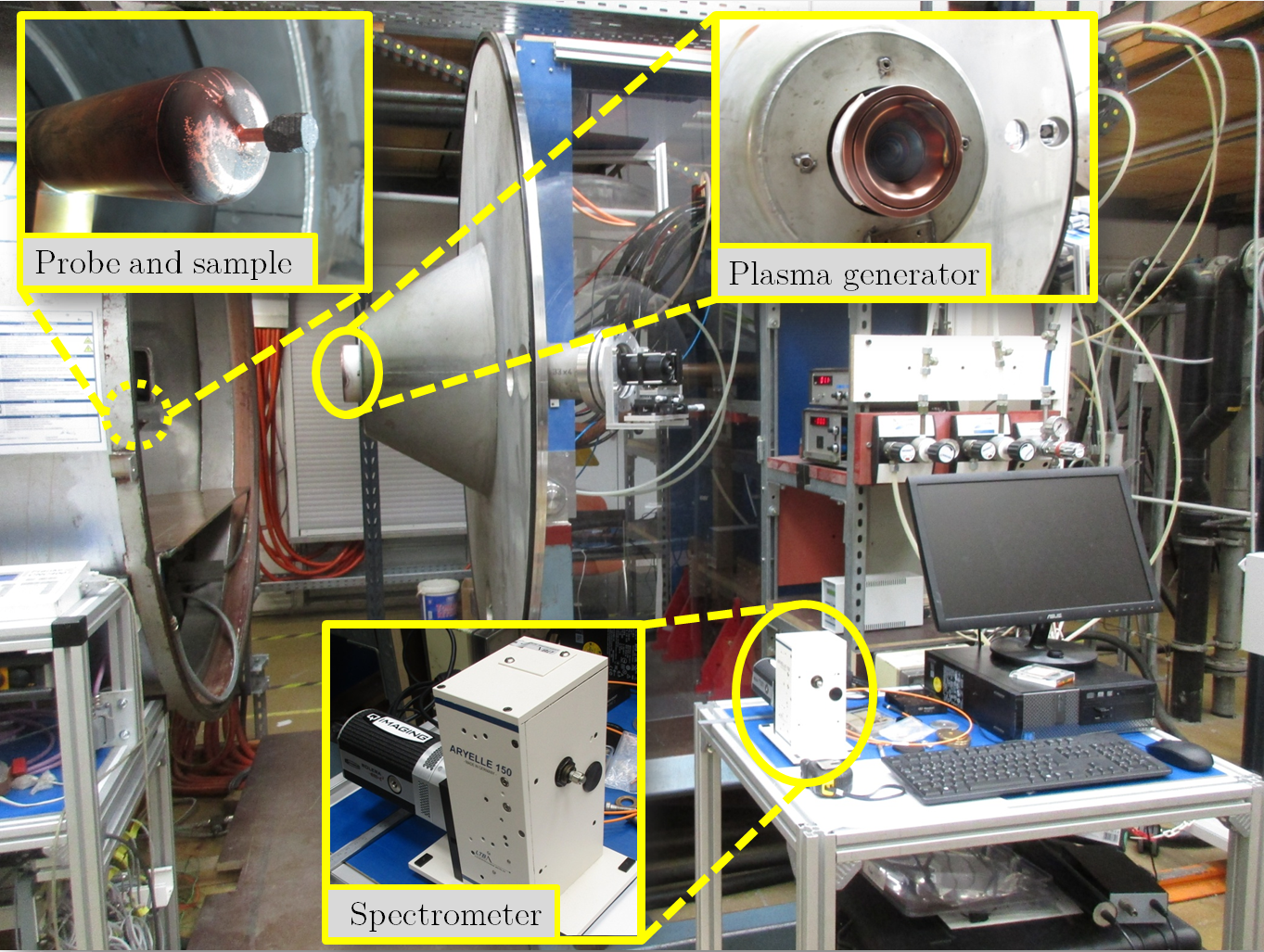}
      \caption{General layout of the facility. Upper panel: The wind tunnel has several windows, so that a variety of instruments can simultaneously record the experiment that is taking place inside the vacuum chamber. Lower panel: View of the vacuum chamber (open) with an emphasis on the sample probe, plasma generator and spectrometer is shown.
              }
         \label{facilities}
   \end{figure}

\noindent
The samples were exposed to a high enthalpy plasma flow (mixture of N$_2$ and O$_2$) in a vacuum chamber simulating the equivalent air friction of an atmospheric entry speed of about 10 $\mathrm{km\,s^{-1}}$ at 80 $\mathrm{km}$ altitude, resulting in the vapourization of the meteoritic material \citep{Loehle17}. We note that meteoroid speeds are usually between $\sim$10 and $\sim$70 $\mathrm{km\,s^{-1}}$ during atmospheric entry \citep{Brown04, Cevolani08} whereas their final speed before free fall is only a few $\mathrm{km\,s^{-1}}$. Thus, the 10 $\mathrm{km\,s^{-1}}$ speed achieved in the plasma wind tunnel falls well in the typical range of meteor speeds.\\

\begin{figure}[h!]
   \centering
   \includegraphics[width=\hsize]{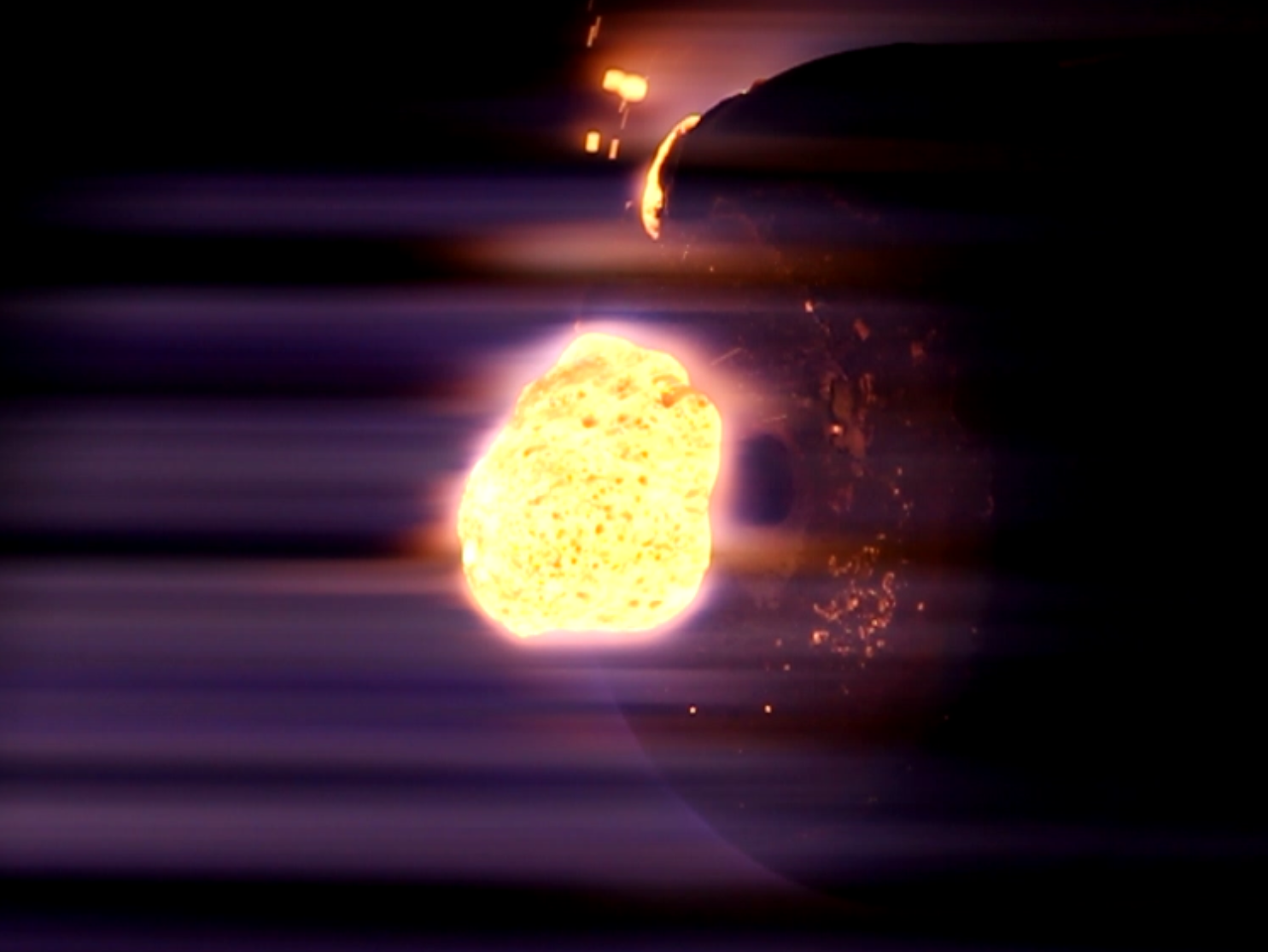}
      \caption{Ablation of Juvinas (eucrite). One can clearly observe the strong effect of the aerothermal heating with the liquid behaviour of the surface flowing along the plasma flow and splashing some droplets of material. In spite of similar experimental conditions and ablation durations ($\sim$2-7 $\mathrm{s}$), measured mass losses are highly variable for the different samples. For example, Agen (H) was significantly less ablated (mass loss of 0.16 $\mathrm{g}$) than Granes (L; mass loss of 2.89 $\mathrm{g}$).}
         \label{Ablation}
   \end{figure}

\begin{figure}[h!]
\centering
  \includegraphics[scale=0.48]{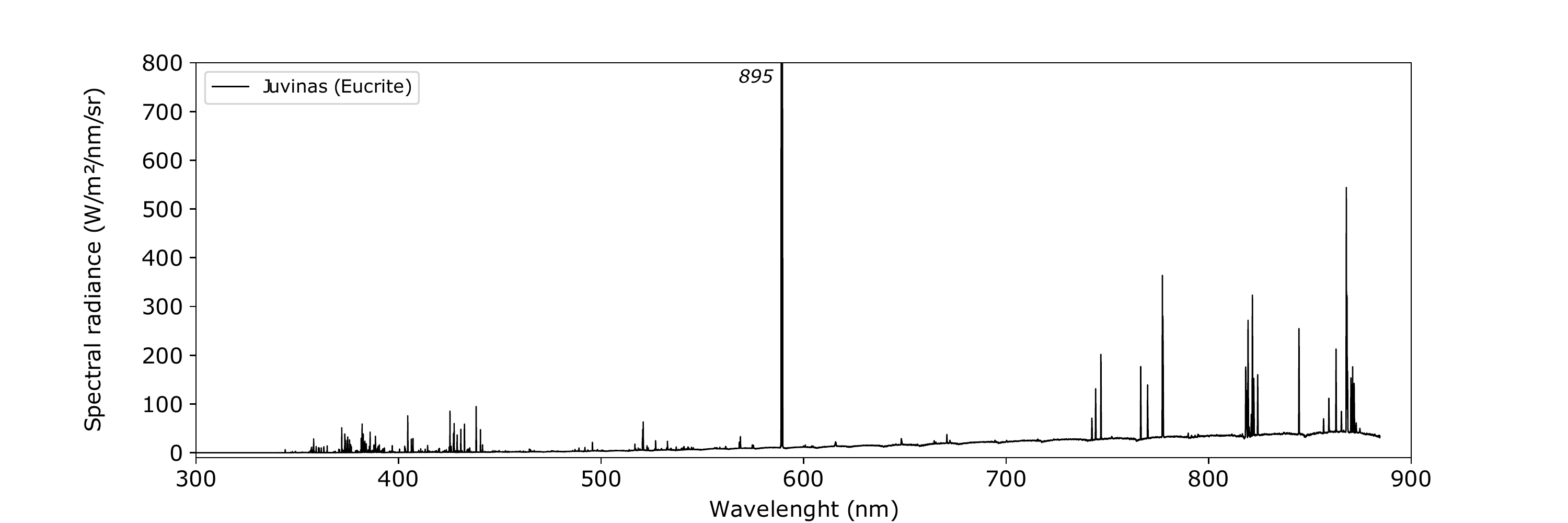}
  \includegraphics[scale=0.48]{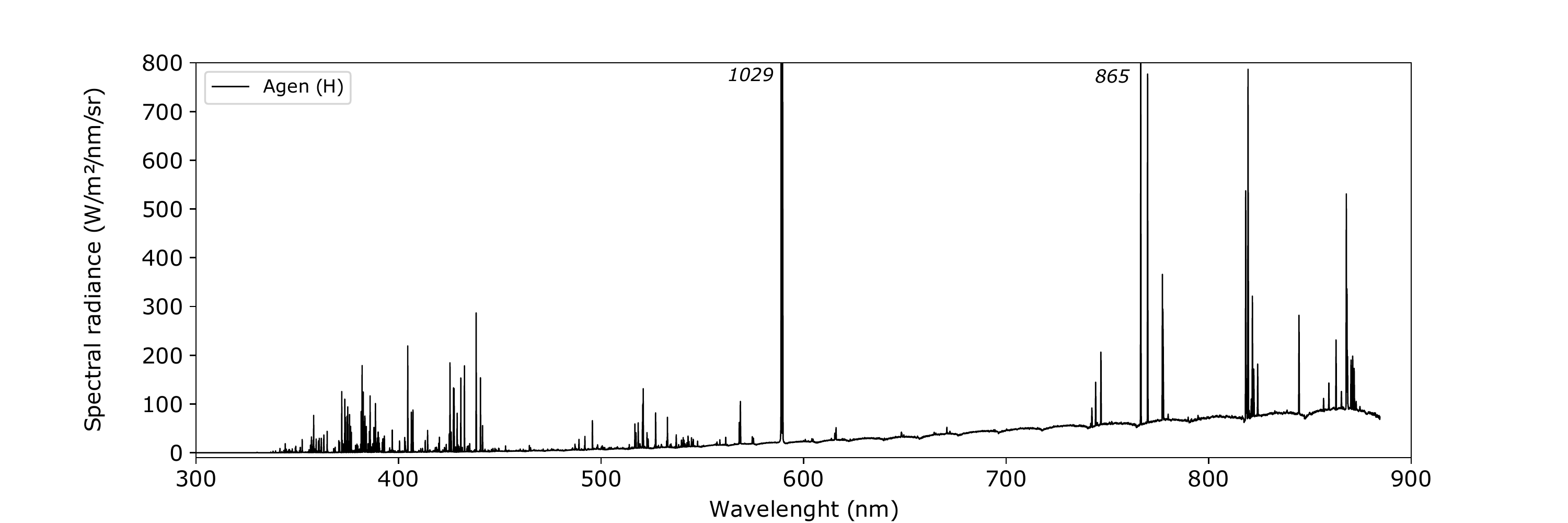}
  \includegraphics[scale=0.48]{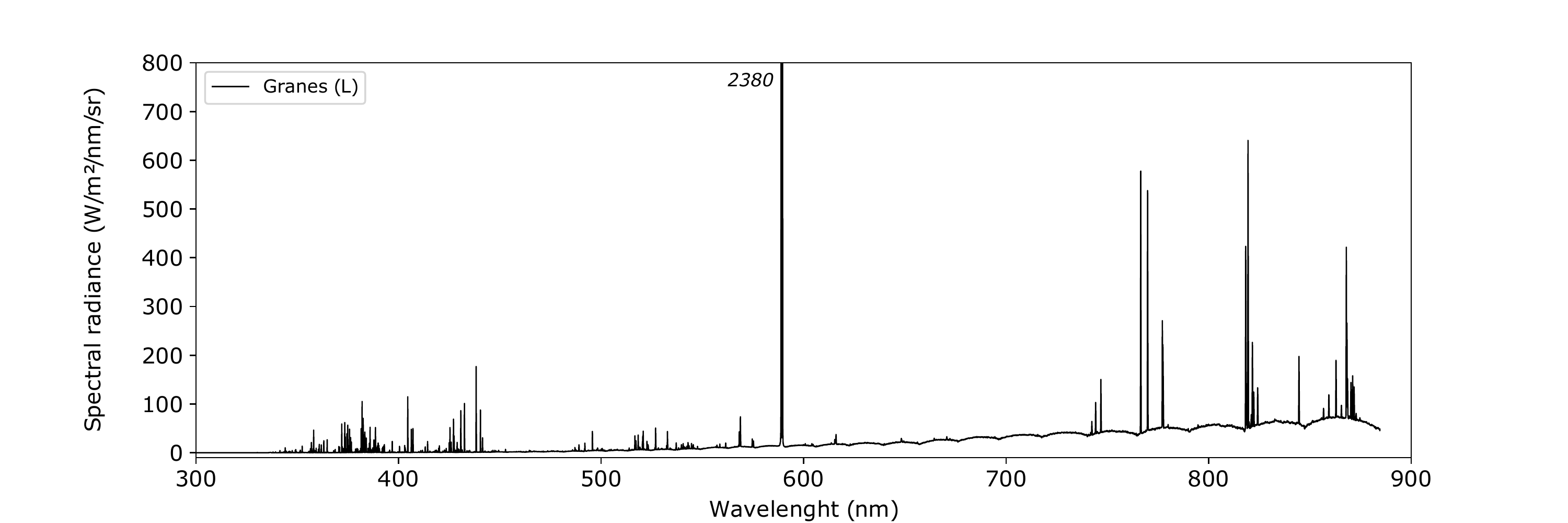}
  \includegraphics[scale=0.48]{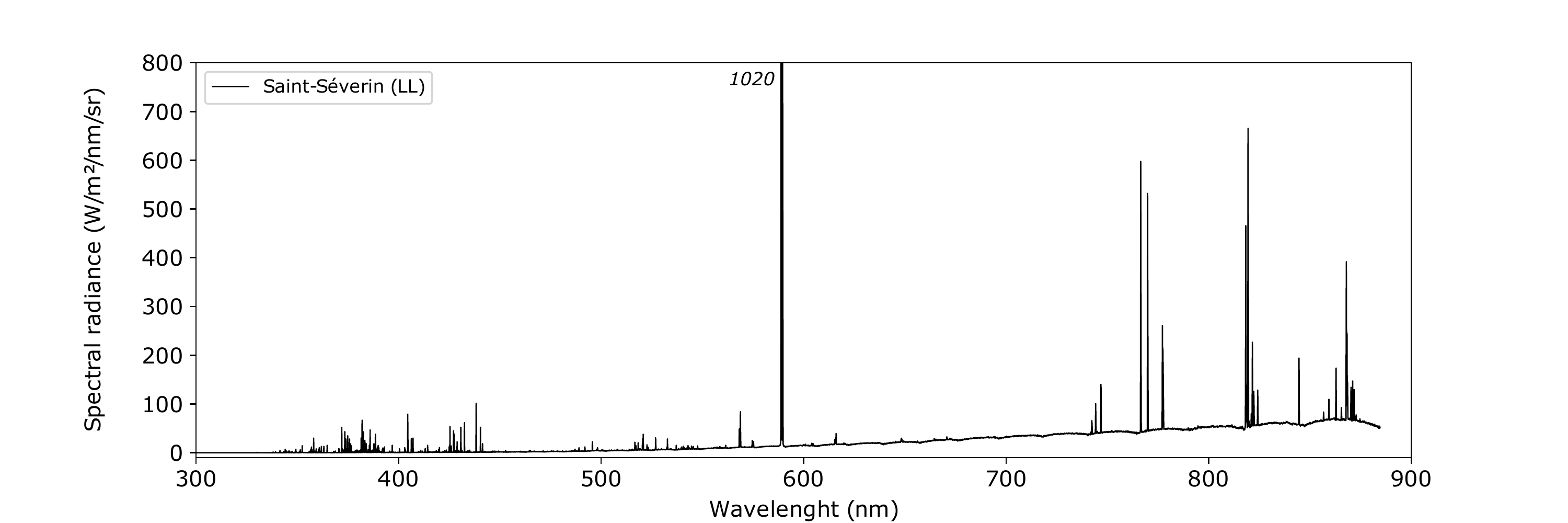}
  \includegraphics[scale=0.48]{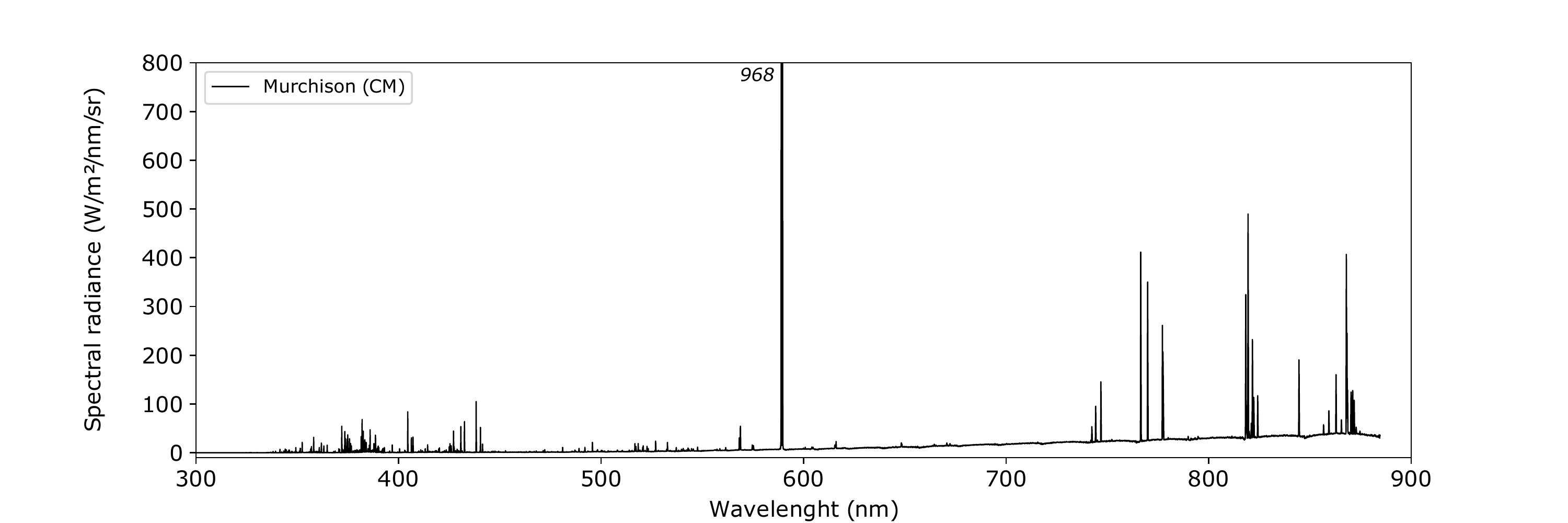}
  \caption{Mean emission spectra of the five meteorite samples.}
  \label{Data}
\end{figure}

\noindent
The general layout of the facility is presented in Fig. \ref{facilities}. Both a camera and a spectrograph, which were both operating in the visible wavelength range, were placed outside the chamber and continuously recorded images (e.g.,  Fig. \ref{Ablation}) and emission spectra (Fig. \ref{Data}), respectively, during the experiments. The video acquisition had a frame rate of 10 $\mathrm{kHz}$. The spectroscopic data were acquired with an Echelle spectrograph (Aryelle 150 of LTB) over the 250-880 $\mathrm{nm}$ wavelength range with an average spectral resolution of $\sim$0.08 $\mathrm{nm\,pix^{-1}}$ \citep{Loehle17} and a frequency of $\sim$10 spectra per second. Several tens to a few hundreds spectra were recorded for each sample during the experiments. For the spectral analysis, we only considered the spectra that were collected during the peak of the emission (which lasted $\sim$2 seconds) that we averaged (average of about 20 spectra per sample) to produce one mean spectrum per sample. The mean spectra of each sample/experiment are presented in Fig. \ref{Data}. The mean spectra recorded by \cite{Loehle17} for the H chondrite EM132, terrestrial argillite, and basalt, respectively, were also used for the present spectral analysis and are shown in Appendix A.

\subsection{Samples}

The samples (see Table \ref{sample}) were selected with the aim of covering as much as possible the meteorite diversity. As a matter of fact, ordinary H, L, and LL chondrites represent $\sim$75\% of the falls and more than 97\% of anhydrous (EH, EL, H, L, and LL) chondrites, CMs are the most common carbonaceous chondrites ($\sim$25\% of all CCs) and eucrites are the most common type of achondrites ($\sim$40\% of all achondrites). These statistics were retrieved from the Meteoritical Bulletin Database\footnote{https://www.lpi.usra.edu/meteor/}. We did not perform an experiment on an iron meteorite as we expect an emission spectrum only formed by metallic (Fe, Ni) lines. Second, the samples (in particular the H, L, and LL suite) were chosen to probe the ability of spectroscopy to distinguish the meteoritic analogues of meteors. Effectively, the elemental composition only slightly varies between the three ordinary chondrite subclasses (see Table \ref{Composition}). 

\begin{table}[h!]
\caption{Samples selected for our experiments with their ablation duration ($\Delta t$).}
\centering
\begin{tabular}{lccc}
\hline
Sample name & Conservation & Clan & $\Delta t$ \\
\hline
Agen       & MNHN &    H    & 2 s \\
Granes     & MNHN &    L    & 7 s \\
St-Séverin & MNHN &    LL   & 3 s \\
Murchison  & NMNH &    CM   & 6 s \\
Juvinas    & MNHN & Eucrite & 7 s \\
\hline
\end{tabular}
\tablefoot{MNHN = Museum National d'Histoire Naturelle, Paris, France. NMNH = National Museum of Natural History, Washington D.C., United States of America.}
\label{sample}
\end{table}

\begin{table}[h!]
\singlespacing
\small
\caption{Mean bulk chemical composition (weight \%) of the various meteorite classes. For Murchison (CM) and Saint-Séverin (LL), the bulk composition was retrieved from the literature \citep{Jarosewich90}. For the H, L, and eucrite meteoritic samples, we used the mean bulk composition of the respective classes from \cite{Hutchison04}. We also add the bulk composition (weight \%) of terrestrial argillite and basalt, measured at the French institute SARM using mass spectrometry. Spectra for the latter two samples were presented in \cite{Loehle17} but analysed in the present paper.}            
\label{Composition}
\centering                        
\begin{tabular}{l c c c c c c c}       
\hline\hline                
 & \small{H} & \small{L} & \small{LL} & \small{CM} & \small{HED} & \small{Argilite} & \small{Basalt} \\    
\hline
&&&\\                        
 Si & 16.9  & 18.5  & 19.0  & 13.3  & 23.0 & 14.66  & 20.17 \\      
 Ti & 0.060 & 0.063 & 0.069 & 0.069 & 0.35 & 0.27   & 1.52  \\
 Al & 1.13  & 1.22  & 1.10  & 1.12  & 6.93 & 5.95   & 6.85  \\
 Cr & 0.366 & 0.388 & 0.41 & 0.303 & 0.22 & 0.0052 & 0.043 \\
 Fe & 27.5  & 21.5  & 19.1  & 20.6  & 14.3 & 2.66   & 8.74  \\ 
 Mn & 0.232 & 0.257 & 0.22 & 0.160 & 0.43 & 0.041  & 0.16  \\      
 Mg & 14.0  & 14.9  & 15.2  & 11.9  & 4.15 & 2.45   & 6.66  \\
 Ca & 1.25  & 1.31  & 1.40  & 1.29  & 7.43 & 15.57  & 7.02  \\
 Na & 0.64  & 0.70  & 0.73  & 0.30  & 0.30 & 0.096  & 2.88  \\
 K  & 0.078 & 0.083 & 0.091 & 0.029 & 0.025& 1.59   & 1.44  \\
 P  & 0.108 & 0.095 & 0.103 & 0.100 &   -  &   -    & 0.39  \\      
 Ni & 1.60  & 1.20  & 0.89  & 1.20  &   -  &0.0024  & 0.0026\\
 Co & 0.081 & 0.059 & 0.040 & 0.058 &   -  &0.00087 & 0.0048\\
 S  & 2.0   & 2.2   & 3.4   & 2.3   & 0.07 &   -    &   -   \\
 H$_2$O& -  &   -   &   -   & 12.6  &   -  &   -    &   -   \\
 C  & 0.11  & 0.09  & 0.022  & 1.88   &   -  &   -    &   -   \\      
 O  & 35.7  & 37.7  & 38.275  & 43.2  & 43.2 &  31.0  & 42.8  \\
   &&&\\
\hline                                   
\end{tabular}
\tablefoot{SARM = Service d'Analyse des Roches et Minéraux, Strasbourg, France.}
\end{table}

   
\onecolumn
\twocolumn   
   
\section{Spectral analysis}

\noindent
 All spectra consist of a thermal continuum (grey body of the sample) on which atomic emission lines (gaseous phase) are superimposed. As a first step (Section 3.1), we identified all emission lines and compared the occurrence of each element, in terms of number of lines with respect to the total number of lines, with the bulk composition (Table \ref{Composition}). As a second step (Section 3.2), we derived mass ratios for each vapourized sample (Mg/Fe and Na/Fe) using two approaches to be directly compared with its bulk composition (Table \ref{Composition}). 

\subsection{Identification and characterization of the emission lines}

\noindent
A first visual inspection of the data reveals an overall spectral similarity from one sample to another, and there are two main groups of lines at short (350-580 $\mathrm{nm}$) and long (700-900 $\mathrm{nm}$) wavelengths, respectively, separated by two very bright sodium lines at 589 $\mathrm{nm}$ (see Fig. \ref{Data}). The only visible difference is the difference in line intensity between the spectra (e.g. the lines of the Agen meteorite are more intense than those of Granes).\\ 

\noindent
As a next step, we identified for all spectra the atomic element associated with each emission line using the NIST database\footnote{https://www.nist.gov/pml/atomic-spectra-database} as well as the total number of lines per element. An example of the line identification step is highlighted for the H chondrite EM132 in Appendix A where 304 lines were identified (a non-exhaustive  list of lines is presented in Appendix B). First of all, it appears that all the identified lines are neutral, that is no ionized lines are present in our spectra. Second, it appears that the first group of lines (350-580 $\mathrm{nm}$ range) comprises only elements that originate from the sample itself [iron (Fe), magnesium (Mg), manganese (Mn), calcium (Ca), sodium (Na), chromium (Cr), potassium (K), hydrogen (H)], whereas the second group of lines (700-900 $\mathrm{nm}$ range) comprises the atomic lines of the plasma radiation [nitrogen (N) lines]. The few oxygen (O) lines in the second group can be attributed to the sample and the plasma but in unknown proportion.\\

\begin{figure} [h!]
   \centering
   \includegraphics[width=\hsize]{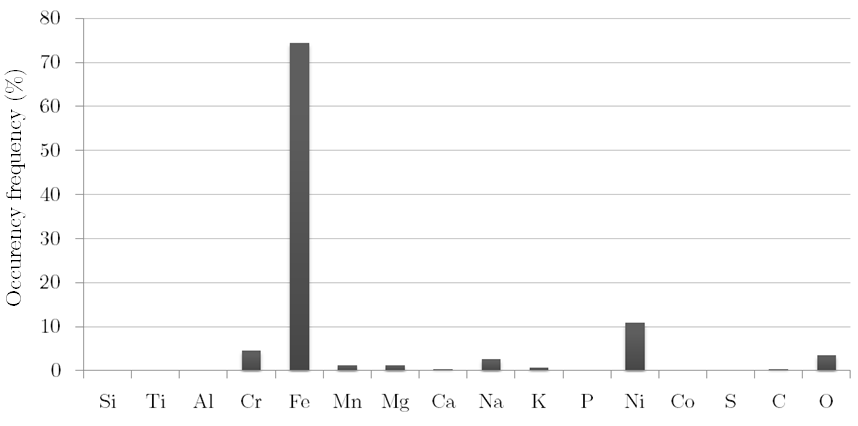}
   \includegraphics[width=\hsize]{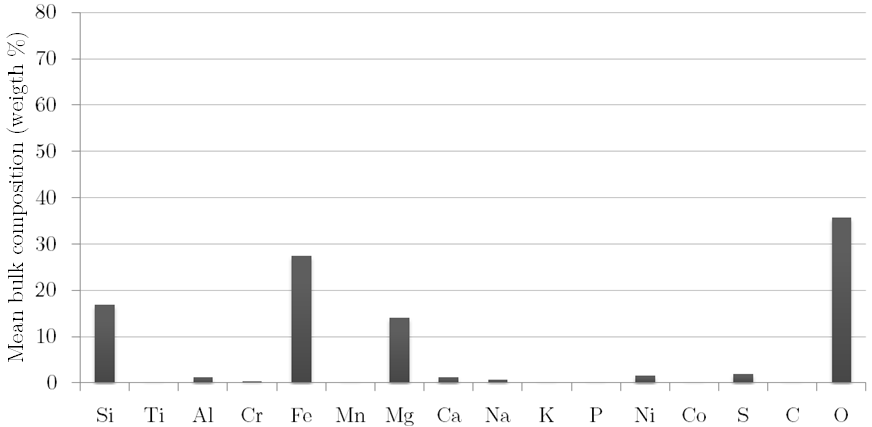}
      \caption{Upper panel: Spectral line distribution per element (\%). Lower panel: Mean bulk composition is shown. The values plotted here come from the $EM132$ spectrum (H chondrite). 
              }
         \label{Line}
   \end{figure}

\noindent
In Fig. \ref{Line} we show for the H chondrite EM132 the frequency of occurrence of lines for each element that we compare with the average bulk composition (weight \%) of H chondrites \citep{Hutchison04}. It appears that the discrepancy between the two histograms is significant. Three of the four most abundant elements in H chondrites (O, Si, Mg) are either missing (Si) or are under-represented (O, Mg). On the other hand, Fe is significantly over-represented (iron lines represent 75\% of all lines). It thus appears that for a given meteorite, the frequency of occurrence of lines for each element does not reflect the sample composition. This result is not surprising and simply reflects the number of emission lines per atom and is in perfect agreement with the spectral lines listed for each element in the NIST database; Fe lines largely dominate the list.\\

\noindent
We obtained similar results for the other meteorite classes except for the eucrite Juvinas whose spectrum is depleted in Ni lines, in agreement with the Ni-free composition of eucrites and achondrites in general \cite{Jarosewich90}. In the case of the two terrestrial samples (argillite and basalt), there are no Fe and Ni emission lines. This, as in the eucrite case, is in perfect agreement with their Fe- and Ni-free bulk composition. 

\noindent
We investigated, as a next step, whether the absolute intensity of the emission lines, without considering the emission coefficient, for different atoms (Na, Mg, Fe, Ni, and Cr) could reflect to some extent their intrinsic abundance. It appears that it is not the case. The two Na emission lines are by far the brightest in all spectra whereas Na is only a minor component (in weight \%) of our samples. The same applies for Cr to a lesser extent. It thus appears that line intensities do not reflect the abundance (in weight \%) of a given element.


\subsection{Abundance ratio trends}

We used the opportunity that the spectra of the various meteorite classes were obtained in similar experimental conditions to compare the trend in mass ratio derived from the spectra to that expected from their mean bulk composition. Specifically, we followed two approaches. Concerning the first case, we tested whether line intensities are directly proportional to the abundance \citep{Nagasawa78}. The second approach consisted in modelling the spectral lines following the auto-absorption model by \cite{Borovicka93} whose output gives the abundance of each element. For the two approaches, we focussed on the following relative abundances (Mg/Fe, Na/Fe) because Mg and Fe are among the most abundant elements in the considered meteorites and because Na, although a minor component, is by far the element with the brightest lines.

\subsubsection{Intensity ratios: Pure emission}

\begin{figure} [h!]
   \centering
     \includegraphics[width=\hsize]{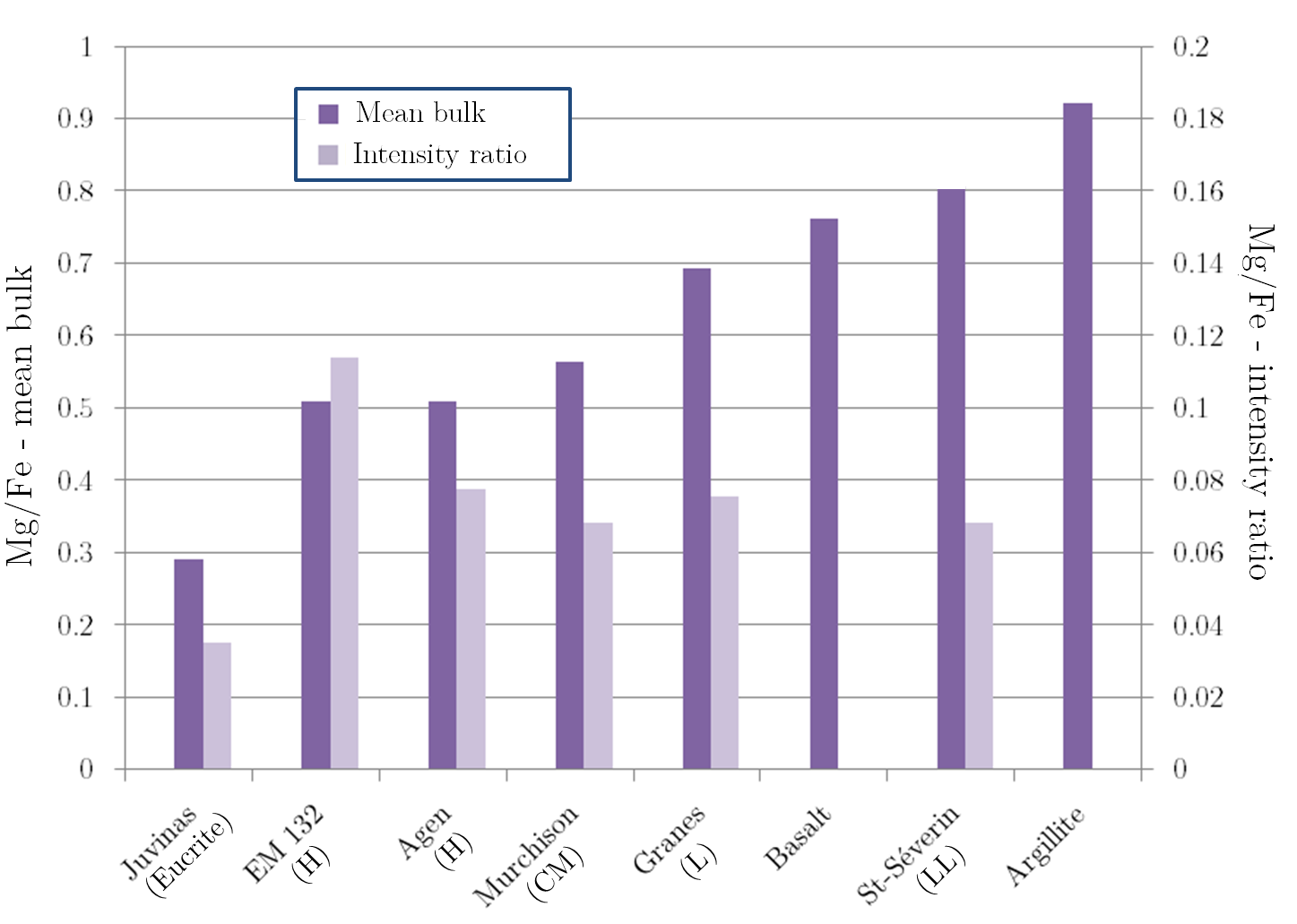}
     \includegraphics[width=\hsize]{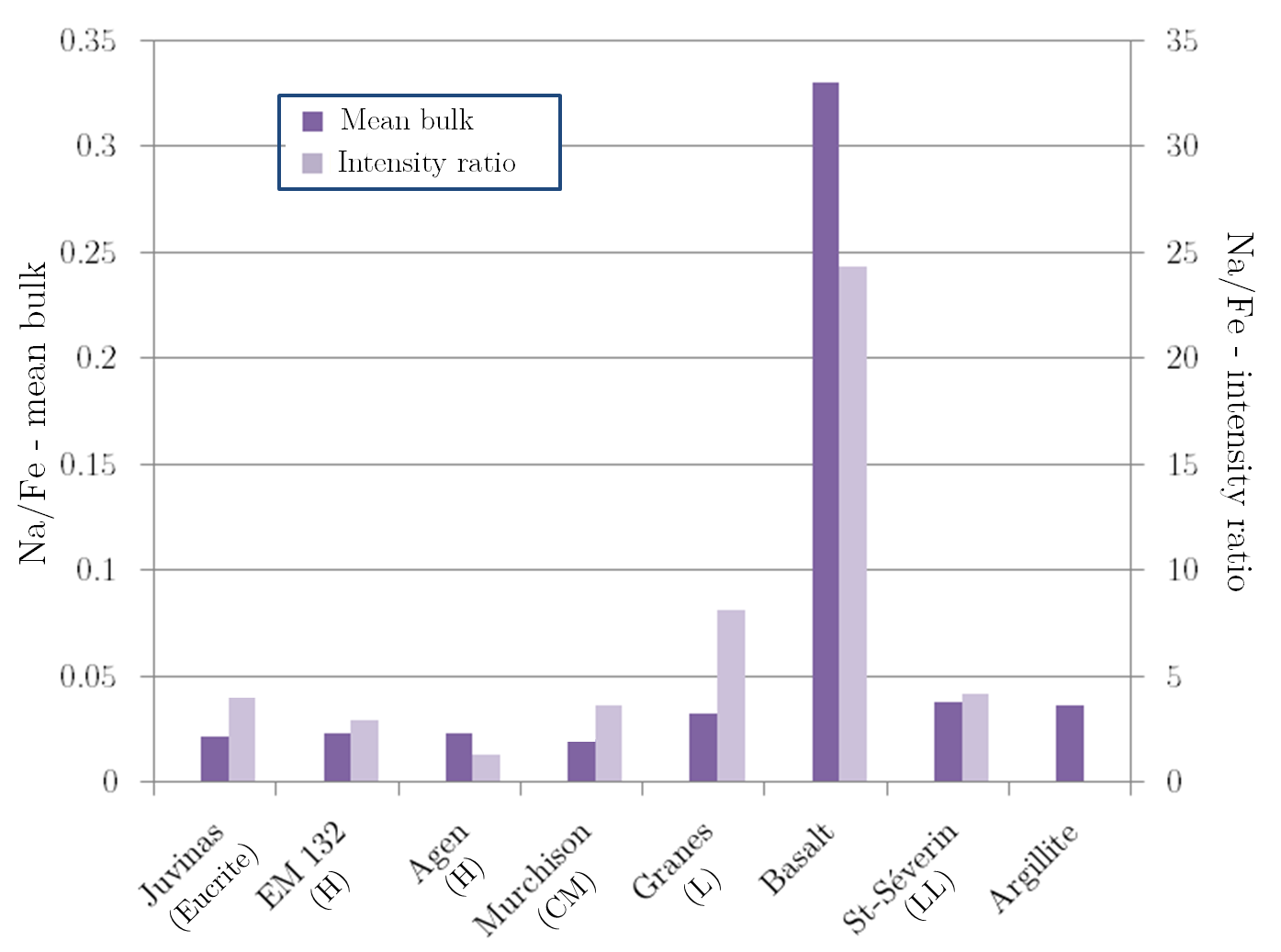}
      \caption{Intensity ratios weighted by the atomic masses and compared to the mass ratios derived from the mean bulk densities (deep purple in both plots). Upper panel: Mg/Fe is shown. Lower panel: Na/Fe is shown. For argillite, no iron was observed so the ratio was not computed. 
              }
         \label{Direct}
   \end{figure}

\noindent
For all spectra, we computed two intensity ratios (Mg/Fe and Na/Fe) using the brightest lines for each element (Fe: 438.35 $\mathrm{nm}$; Mg: 518.36 $\mathrm{nm}$; and Na: 589.00 $\mathrm{nm}$) and compared the trend to that of their mean bulk composition. Such an approach is valid in the case of pure emission \cite{Nagasawa78}, because in that case the intensity of a given line is proportional to the atomic abundance of the emitting element (see Appendix B for more details). The results are shown in Fig. \ref{Direct}. It appears that the compositional trend from one meteorite to the other is rarely respected (especially in the case of Mg/Fe; see for example H versus LL). It thus appears that intensity ratios are not informative of the relative abundance (weight \%) of given atomic species and cannot be used to determine the meteoritic analogue. This may stem from the fact that we neglect the auto-absorption of the gas and its temperature. A more sophisticated model, such as that presented in the next section may thus improve those results. 

\subsubsection{Spectral modelling: Autoabsorption model}

\begin{figure} [h!]
   \centering
   \includegraphics[width=\hsize]{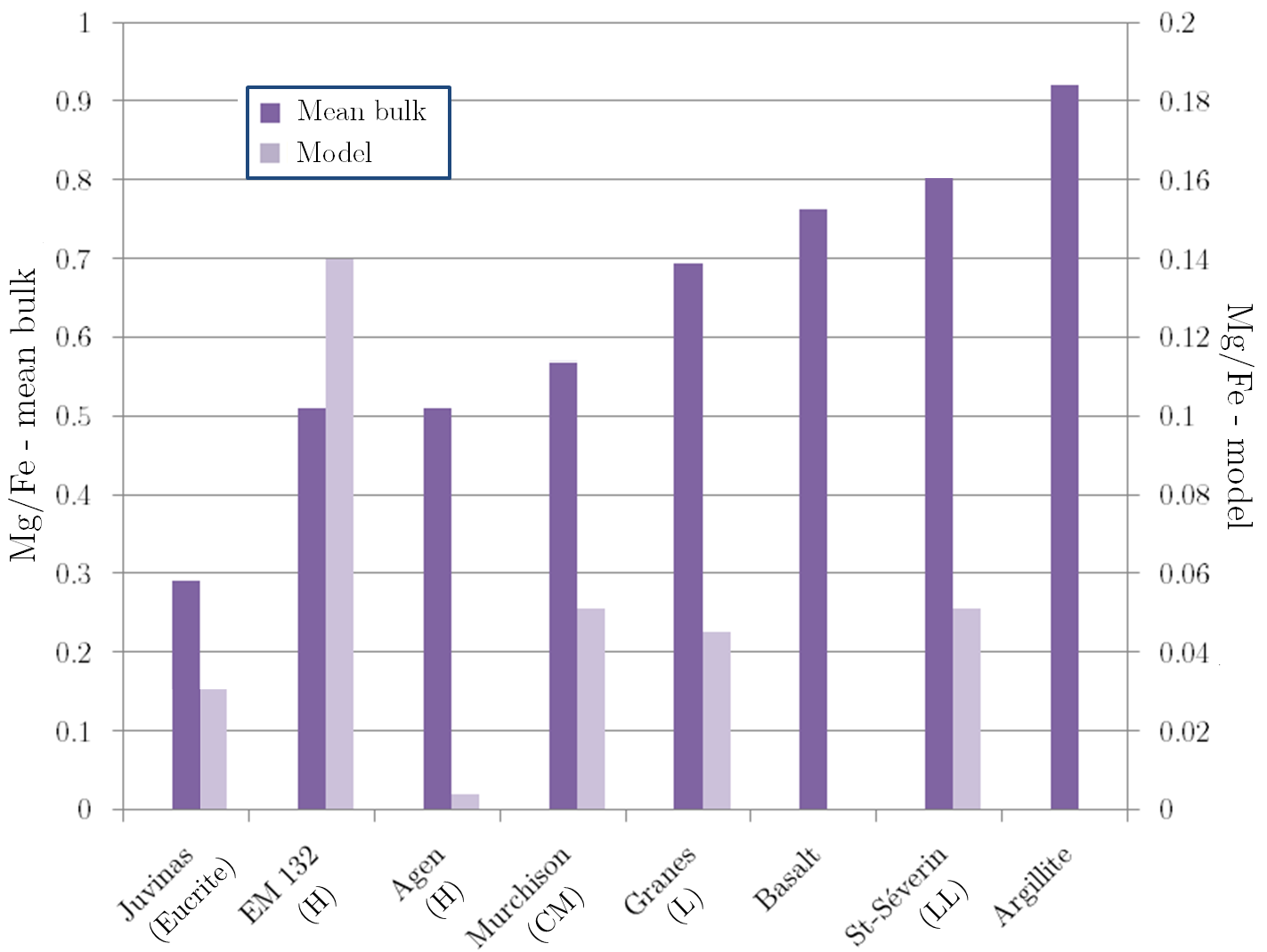}
   \includegraphics[width=\hsize]{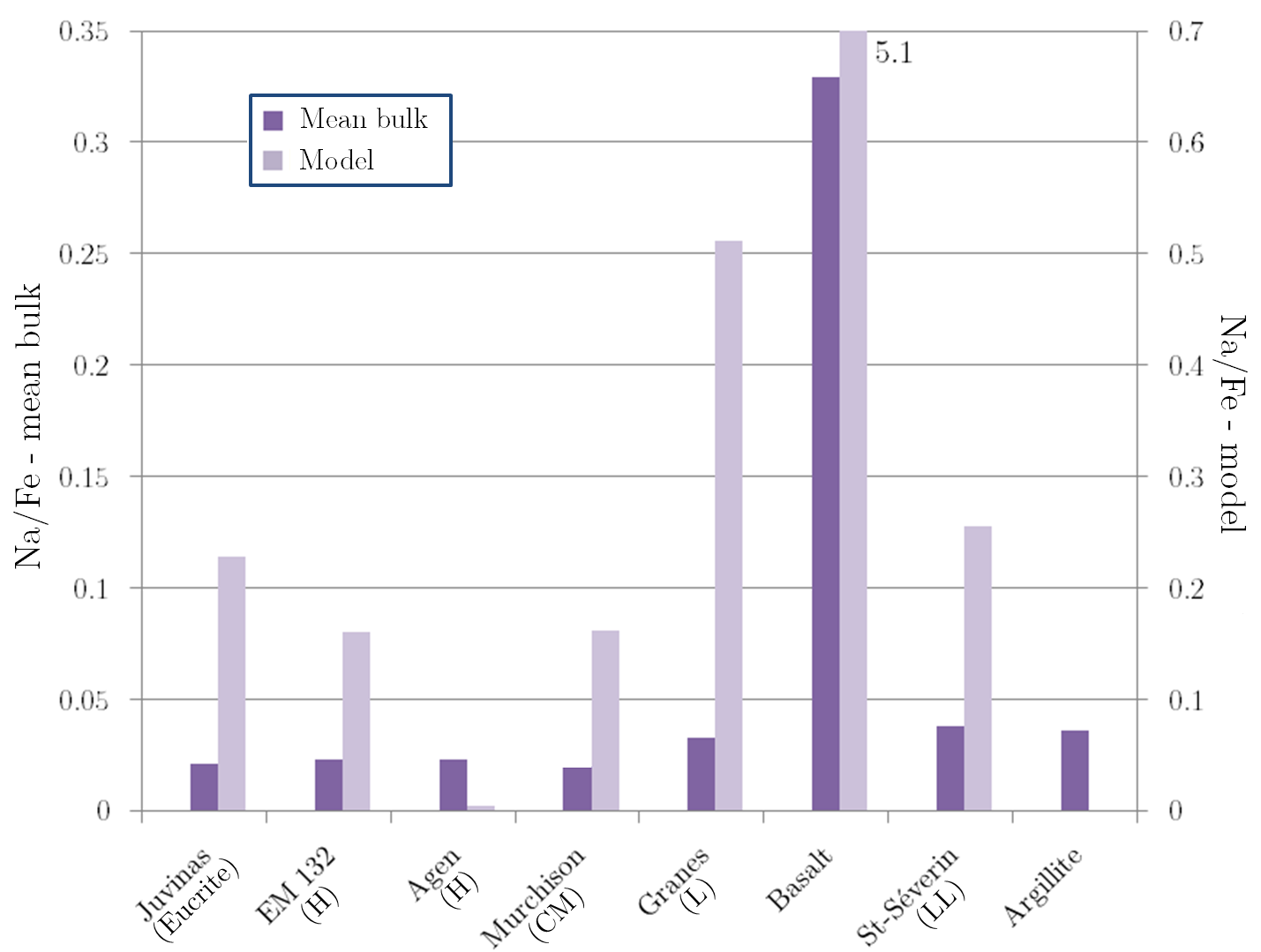}
      \caption{Mass ratios derived from the model compared to the average composition of the sample. For argillite, no iron was observed so ratios were not computed. Upper panel: The derived Mg/Fe is shown in light purple and the expected one in deep purple. Lower panel: The derived Na/Fe is shown in light purple and the expected one in deep purple.
              }
         \label{Model}
   \end{figure}

\noindent
Our method is based on the auto-absorption model by \cite{Borovicka93} and has only two free parameters, namely the gas temperature and the column density  (see Appendix B for a description of the model). Both variables were constrained by minimizing the difference between the integrated area under the spectral line and model. Specifically, we first constrained the temperature of the gas by applying the model to the iron lines. This allowed us to derive a gas temperature of about 4000 $\mathrm{K}$. Such mean temperature value was obtained by best-fitting the iron lines assuming a similar column density. We then applied the model to other spectral lines with a fixed temperature of 4000 $\mathrm{K}$ to constrain the column densities of each element; we made this assumption of applying the model to all the spectral lines because the experimental conditions were the same for all samples. The derived column densities are presented in Table \ref{results_densities}. We also derived the surface temperature of each sample by fitting the thermal continuum of its spectrum with a Planck function (emissivity of $0.83$ following \cite{Loehle17}). It appears that all surface temperatures fall in the 2100-2500 $\mathrm{K}$ range. Finally, we derived the Mg/Fe and Na/Fe mass ratios for all samples and compared them with those expected from their mean bulk composition (Fig. \ref{Model}).

\begin{table}[h!]
\caption{Atomic column densities derived from the output column densities for sodium (line at 589.00 $\mathrm{nm}$), magnesium (line at 518.36 $\mathrm{nm}$) and iron (line at 438.35, 526.95, and 532.80 $\mathrm{nm}$)).}
\centering
\begin{tabular}{lccc}
\hline
Sample & $N_{Na}$ & $N_{Fe}$ & $N_{Mg}$ \\
 &($10^{15} cm^{-2}$) & ($10^{15} cm^{-2}$) & ($10^{15} cm^{-2}$) \\
\hline
Agen       & 6.02 & 484.0  & 3.99  \\
Argilitte  & 0.22 & -     & -     \\
Basalt     & 3.79  & 0.31  & -     \\
EM132      & 1.90  & 4.84  & 1.59  \\
Granes     & 19.0  & 15.3  & 1.59 \\
Juvinas    & 1.35  & 2.42 & 0.16     \\
Murchison  & 3.80  & 9.66 & 1.12 \\
St-Séverin & 6.02 & 9.66 & 1.12 \\
\hline
\end{tabular}
\label{results_densities}
\end{table}

\noindent
Concerning the derived Mg/Fe and Na/Fe mass ratios, we observed - similar to the previous subsection - a significant discrepancy between the inferred and expected mass ratios (a factor of 10 or more). We also observed that for a similar bulk compositions (H, L, and LL chondrites), the compositional trend (increasing Mg/Fe ratios from H to LL) is not retrieved with the L chondrite Granes having the highest mass ratio. Importantly, the two experiments for H chondrites (EM132 and Agen) highlight that the derived uncertainty (with our model) for the Mg/Fe and Na/Fe mass ratios for a given meteorite class is comparable to the expected difference between different meteorite classes (e.g. between H and CM chondrites). Our results therefore imply that there is no straightforward link between the bulk composition of a sample and the spectrum resulting from its ablation.
 
\section{Discussion}

We performed ablation experiments in a wind tunnel of five different meteorite types (H, L, LL, CM, and eucrite) with the specific aim to determine whether spectroscopy can be a useful tool for determining the meteoritic analogue of meteors. A spectral analysis of the obtained spectra reveals a number of difficulties that we summarize hereafter:\\

\begin{enumerate}[(i)]

\item Silicon (Si) is one of the four major atomic elements (in weight \%) present in all five meteorites. Yet, the emission lines associated with this element are absent from all spectra. 
\item Volatiles (e.g. Na) are overestimated with respect to other elements (e.g. Fe or Mg) even though these are minor components.
\item Mass ratios (Mg/Fe and Na/Fe) derived under different assumptions (pure emission and auto-absorption model) do not reflect the true composition of the samples nor relative differences in composition. 

\end{enumerate}

\noindent
Case (i) may be because the temperatures reached during the experiments ($\sim$4000 $\mathrm{K}$) were too low to dissociate the very stable gaseous SiO molecule resulting from the ablation of the solid SiO$_2$ phase and that there were likely no free Si atoms in the gas. Such low experimental temperatures may also explain why all emission lines observed in our spectra are neutral lines. In contrast, meteor spectra reveal the presence of numerous ionized lines (so-called second component at 10000 $K$; see \cite{Borovicka93}), which can be explained by the higher entry velocities ($\sim$20-70 $\mathrm{km\,s^{-1}}$) of bolides with respect to those we simulated in our experiments  ($\sim$10 $\mathrm{km\,s^{-1}}$). Table \ref{Thermo} illustrates this point well by showing the energy required for ionization for each element.\\

\noindent
Case (ii) can be explained, at first order, as the byproduct of two physical quantities, namely the melting and vapourization temperature of a given element (Table \ref{Thermo}). Sodium (Na) and potassium (K) illustrate this case particularly well. First, their favourably high emission coefficient implies that the intrinsic intensity of their emission line is higher than that of other elements with lower emission coefficients. Second, their low fusion and vapourization temperature implies that the gas gets enriched significantly more in those two elements than in others such as iron. As a matter of fact, the high temporal spectra resolution of our measurements allowed us to observe the rapid apparition of volatiles such as sodium with respect to metallic elements such as Fe that appeared later. The temporal shift (about 0.5 $\mathrm{s}$) remains small compared to the duration of the emission peak (about 2 $\mathrm{s}$).\\

\noindent
Case (iii) is a direct implication of (ii). This is particularly true for the first case (pure emission) but should be less true when using the model as the latter takes into account the absorption coefficient. An obvious
caveat of the model is that most elements possess only a few emission lines (e.g. Mg) limiting the ability to have a statistical approach in our analysis. In this respect, only iron and nickel possess enough emission lines (see \ref{Line}).\\ 

\noindent
In summary, the difference in terms of thermodynamical behaviour among the elements leads to a spectral variability among different samples (in terms of intensity of the emission lines) that is very difficult to decode. Such spectral difference can even be observed in the case of samples with nearly identical compositions (see the different values for the two H chondrites; e.g. Figs. \ref{Direct} and \ref{Model}).\\

\noindent
The line identification step however reveals that spectroscopy is able to distinguish samples that are not made of the same elements/atoms. For example, there is no nickel in eucrites and this is well verified via our experiments that show an absence of nickel lines in the Juvinas spectrum. Similarly, we can identify terrestrial samples (argillite and basalt) via the absence of iron and nickel emission lines in their spectra. This has global implication for establishing the ability of spectroscopy to determine the meteoritic analogues of meteors. Whereas all chondrites are made of the same atoms, the same is not true for achondrites and iron meteorites. The latter are made of Fe, Ni, and Co mainly with traces of Ga, Ge and Ir \citep{Jarosewich90}, whereas the former do not have any nickel but do have most elements seen in chondrites (see Table \ref{Composition}). As such, a spectral analysis consisting of a simple line identification should allow distinguishing between chondrites, achondrites, and iron meteorites. For this to be the case, the spectral resolution should be sufficient to distinguish the individual atomic lines ($\sim$0.1 $\mathrm{nm\,pix^{-1}}$). We would like to stress that our findings are fully consistent with the review by \cite{Borovicka2015} in which they propose that meteor spectroscopy can be used to distinguish the main types of incoming meteoroids (chondritic, achondritic, and metallic).

\begin{table}[h!]
\caption{Some physical data of the main important elements of meteorites compositions}             
\label{Thermo}  
\centering                        
\begin{tabular}{lcccc}
\hline
\hline
 & \small{Sublimation} & \small{Fusion} & \small{Vaporization} & \small{1$^{st}$ ionization}\\
 & \small{enthalpy} & \small{temperature} & \small{temperature} & \small{potential}\\
 & \small{($\mathrm{kJ\,mol^{-1}}$)} & \small{($\mathrm{K}$)} & \small{($\mathrm{K}$)} & \small{($\mathrm{eV}$)}\\
\hline
 &&&&\\
H  &   218.00 &   - &   - & 13.6 \\
C  & 716.68 & - & 4100 & 11.3 \\
N  &   472.68 &   - &   - & 14.5 \\
O  &   249.18 &   - &   - & 13.6 \\
Na &  107.30 &  - & 1171 &  5.1 \\
Mg & 147.10 &  923 & 1366 &  7.6 \\
Al & 329.70 &  933 & 2791 &  6.0 \\
Si & 450.00 & 1685 & 3505 &  8.2 \\
K  &  89.00 &  336 & 1040 &  4.3 \\
Ca & 177.8 & 1115 & 1774 &  6.1 \\
Cr & 397.48 & 2130 & 2952 &  6.8 \\
Mn & 283.26 & 1519 & 2235 &  7.4 \\
Fe & 415.47 & 1809 & 3133 &  7.9 \\
&&&&\\
\hline
\end{tabular}
\tablefoot{Data from the NIST database. Thermodynamical parameter: http://webbook.nist.gov/chemistry/form-ser/ and ionization potentials: https://www.nist.gov/pml/atomic-spectra-database.}
\end{table}

\section{Conclusions}

Several camera networks have been installed around the world and have been witnessing the atmospheric entry of several bolides each year. From those datasets, the orbits of the bolides are derived along with the distribution ellipse (location of the fall). However, the vast majority of the bolides yield meteorites in areas unfavourable for their recovery, or are not found on the ground. For those events, although we have a precise orbit determination, we are unfortunately not able to associate a given meteorite group with the orbit. Our objective in this work was to test whether spectroscopy in the visible wavelength range of these bolides could allow us to classify the incoming meteoritic material. To test this hypothesis, we performed ablation experiments in a wind tunnel on five meteorite types (H, L, LL, CM, and eucrite) simulating the atmospheric entry conditions of a meteoroid at 80 $\mathrm{km}$ of altitude for an entry speed of 10 $\mathrm{km\,s^{-1}}$. Visible spectra with a resolution of $\sim$0.1 $\mathrm{nm}$ per pixel were collected during the entire ablation sequence allowing us to measure the composition of the vapourized material.\\ 

\noindent
A detailed spectral analysis including a systematic line identification and mass ratio determinations (Mg/Fe, Na/Fe) with or without a spectral model was subsequently performed on all spectra. It appears that spectroscopy, via a simple line identification, can allow us to distinguish the three main meteorite classes (chondrites, achondrites, and irons), but it has not the potential to distinguish for example an H chondrite from a CM chondrite. It also appears that mass ratios are not informative about the nature of the vapourized material.\\ 

\noindent
Regarding the future of meteor spectroscopy, distinguishing the three main classes of meteorites (chondrites, achondrites, and irons), in particular the fraction of iron-like bolides could be an interesting project as it could help quantify the amplitude of the atmospheric bias with respect to meteorite fall statistics. To investigate such question, spectroscopy in the 400-600 $\mathrm{nm}$ wavelength range with a resolution of 0.1 $\mathrm{nm\,pix^{-1}}$ would be ideal as it would allow us to resolve the emission lines of the key elements (Fe, Ni, Mg, Cr, Na, and Si). \\

\noindent
To conclude, the source location within the main belt of the different meteorite classes (H, L, LL, CM, CI, $etc$.) should continue to be investigated via fireball observation networks. Spectroscopy of incoming bolides will only marginally help classify the incoming material precisely (iron meteorites only). To reach a statistically significant sample of recovered meteorites along with accurate orbits (>100) within a reasonable time frame (10-20 years), the optimal solution maybe to extend spatially existing fireball observation networks.


\begin{acknowledgements} 
We thank the Programme National de Plan\' etologie, the Laboratoire d'Astrophysique de Marseille and the Agence Nationale de la Recherche (FRIPON project: ANR-13-BS05-0009) for providing financial support for these experiments. 
\end{acknowledgements}


\bibliographystyle{aa}

\onecolumn

\Online
\begin{appendix}
\section{Additional figures}
\begin{figure}[h!]
\centering
  \includegraphics[scale=1]{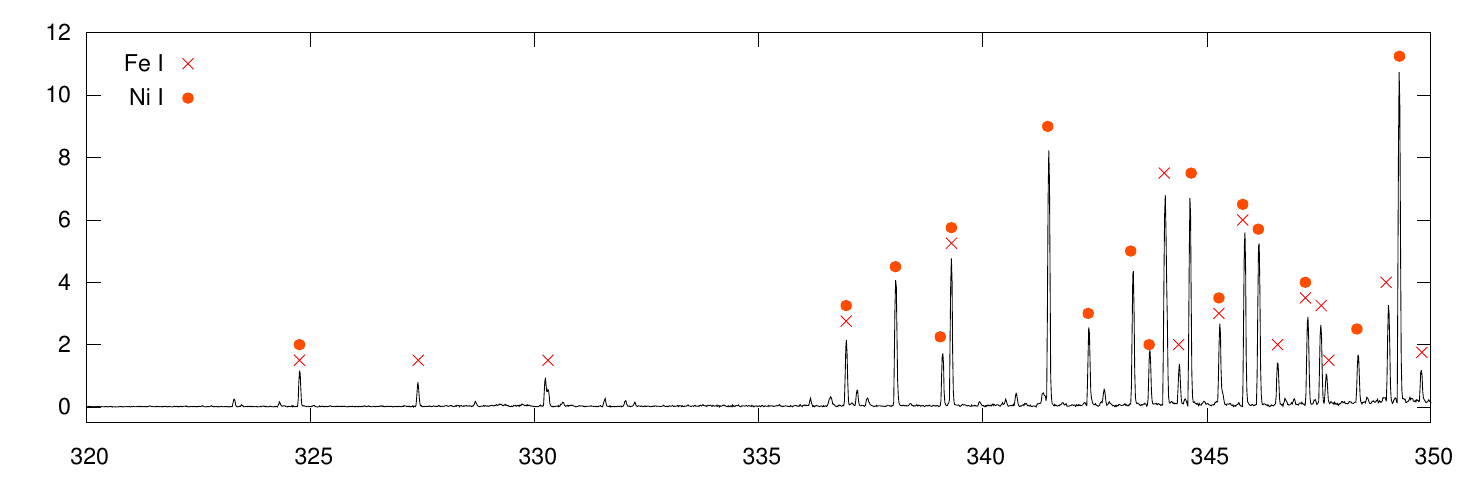}
  \includegraphics[scale=1]{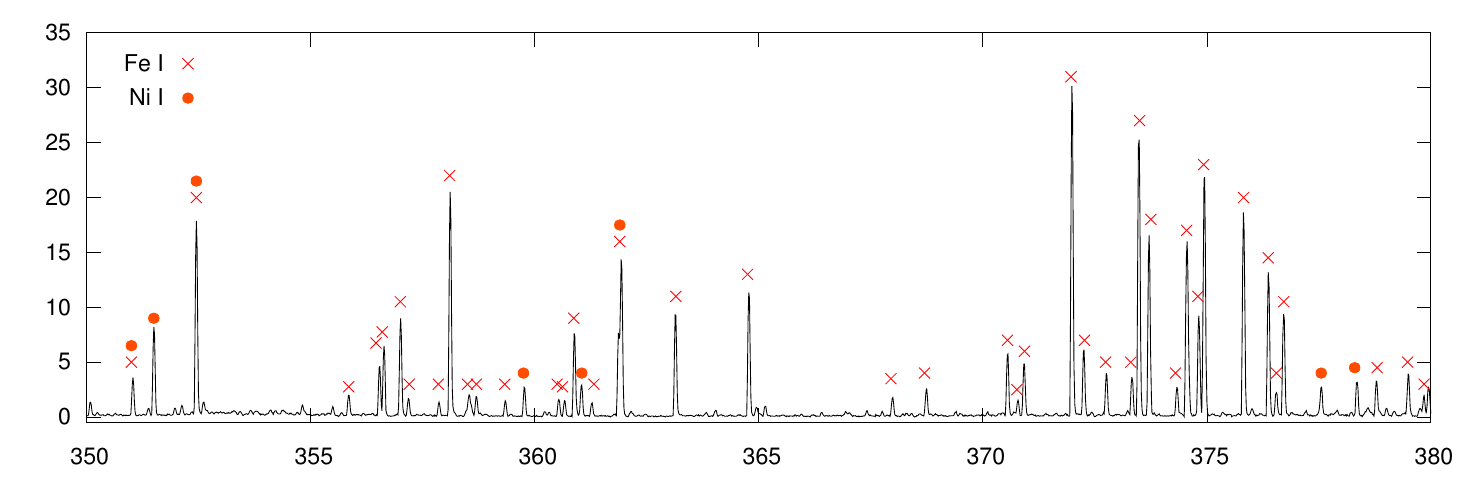}
  \includegraphics[scale=1]{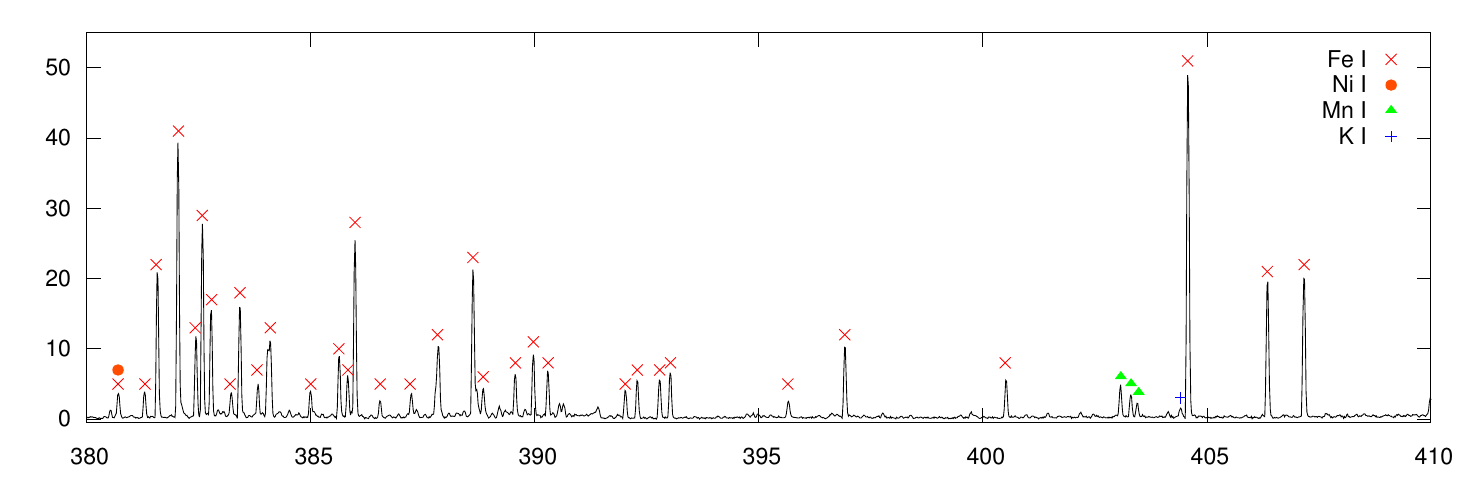}
  \includegraphics[scale=1]{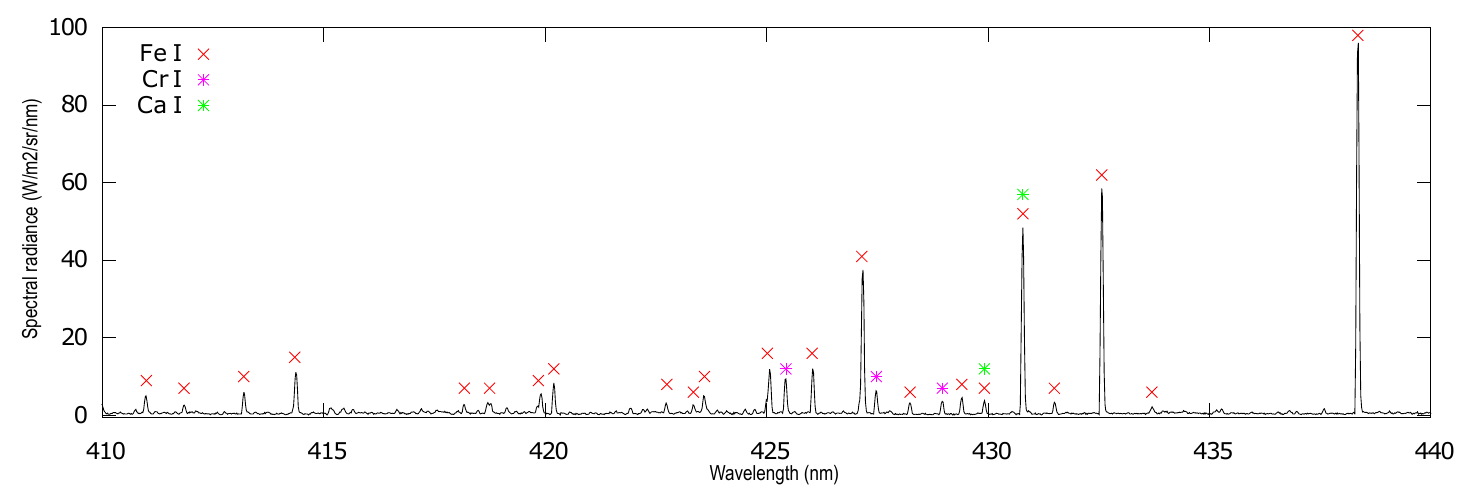}
  \caption{Identification of the emission lines of the EM132 spectrum (H chondrite) using the NIST database (shift <0.1 $\mathrm{nm}$).}
  \label{spectrum}
\end{figure}

\begin{figure}[h!]
\centering
  \includegraphics[scale=1]{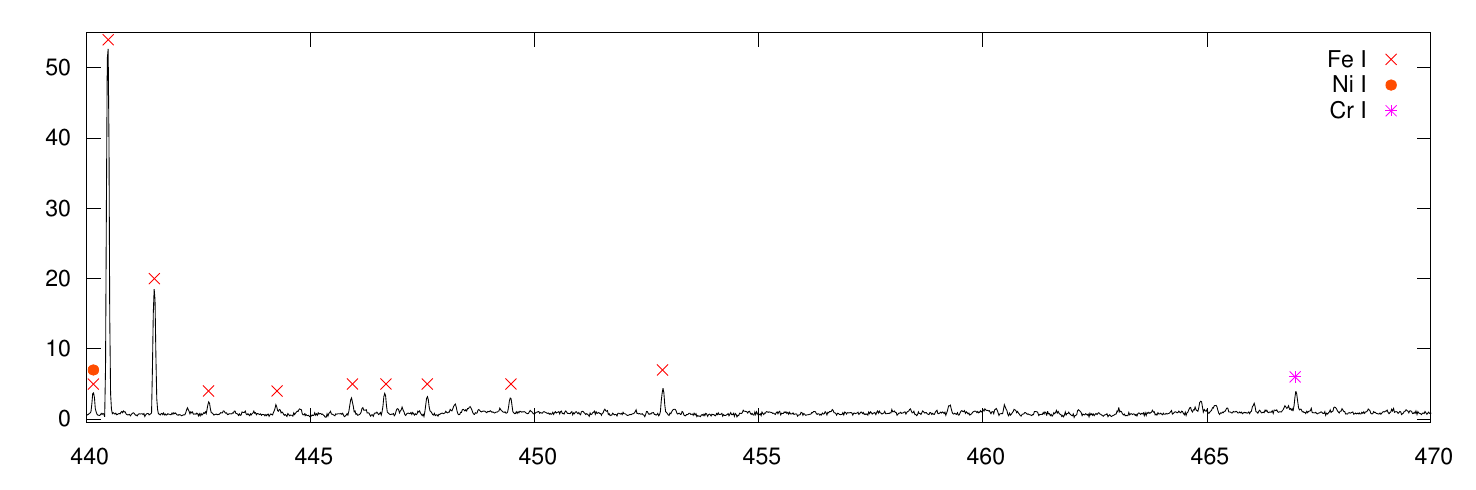}
  \includegraphics[scale=1]{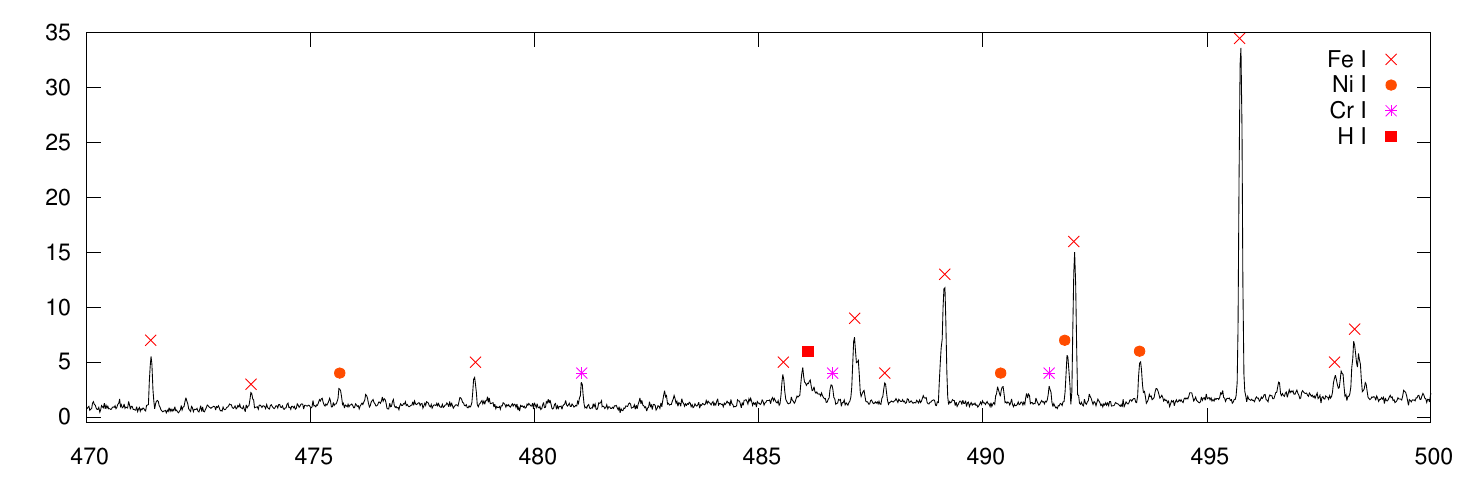}
  \includegraphics[scale=1]{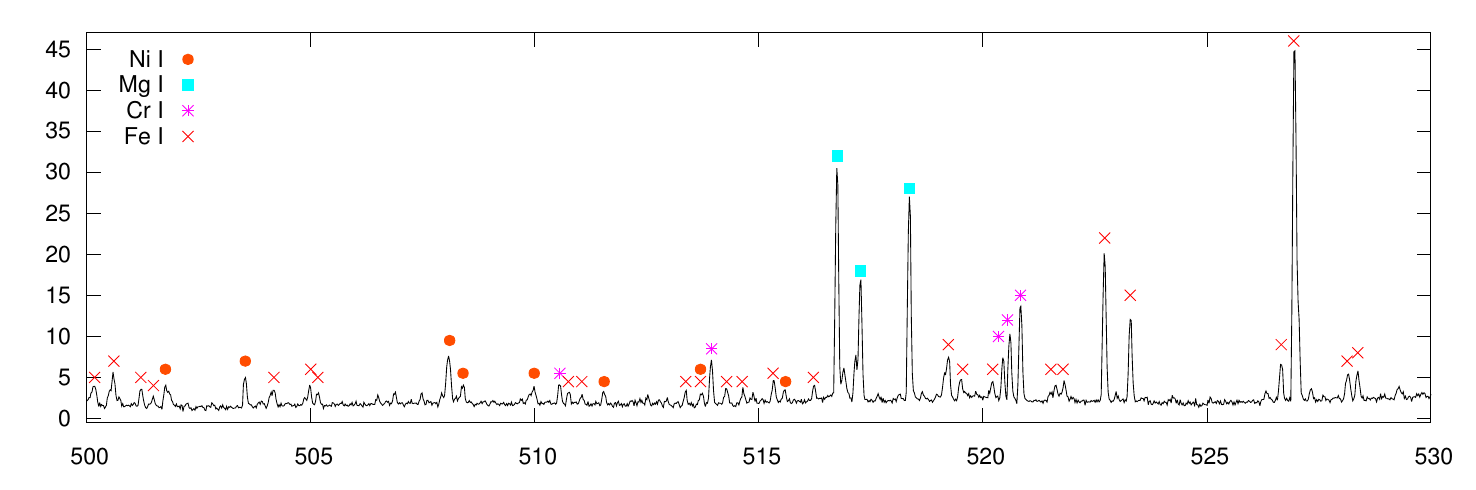}
  \includegraphics[scale=1]{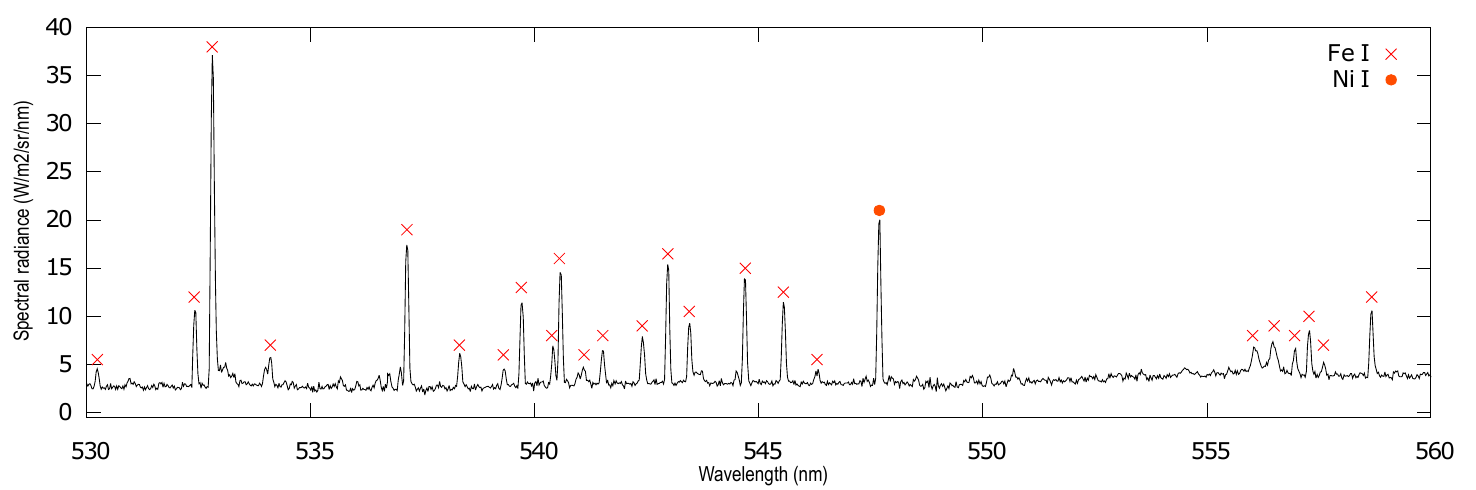}
  \captionsetup{labelformat=empty}
  \caption{Fig A.1 (\emph{Continued})}
\end{figure}

\begin{figure}[h!]
\centering
  \includegraphics[scale=1]{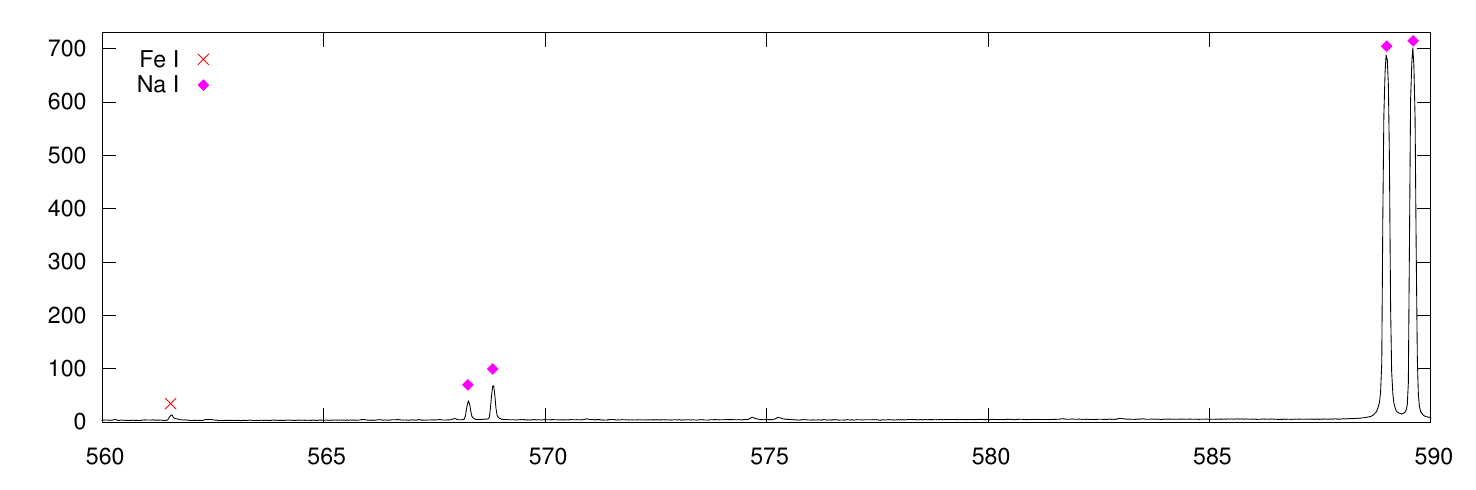}
  \includegraphics[scale=1]{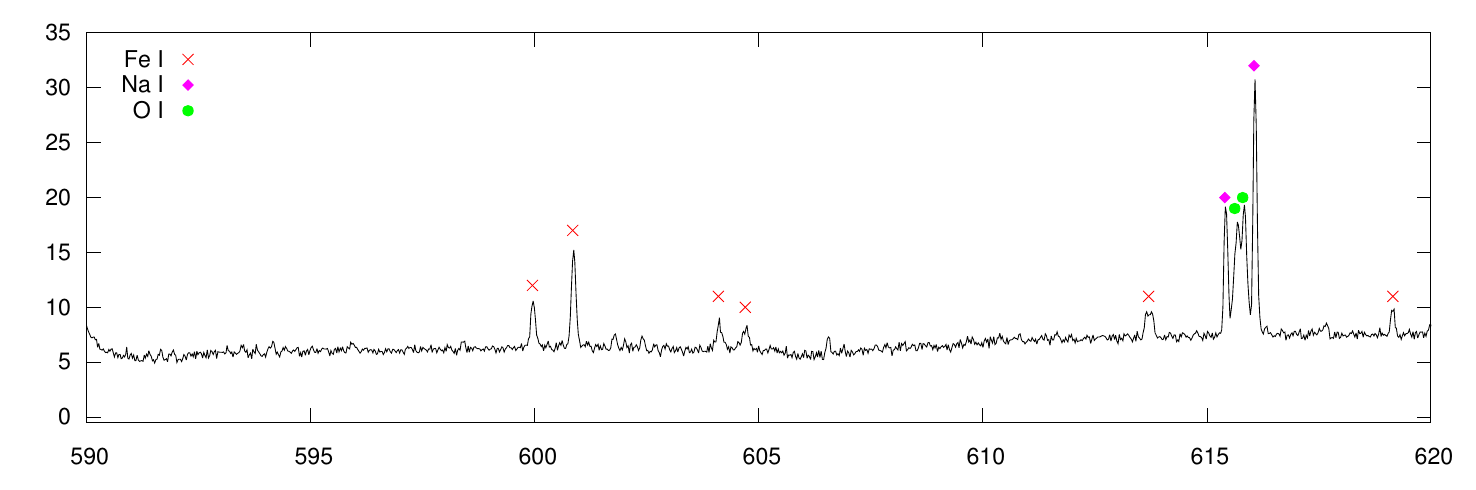}
  \includegraphics[scale=1]{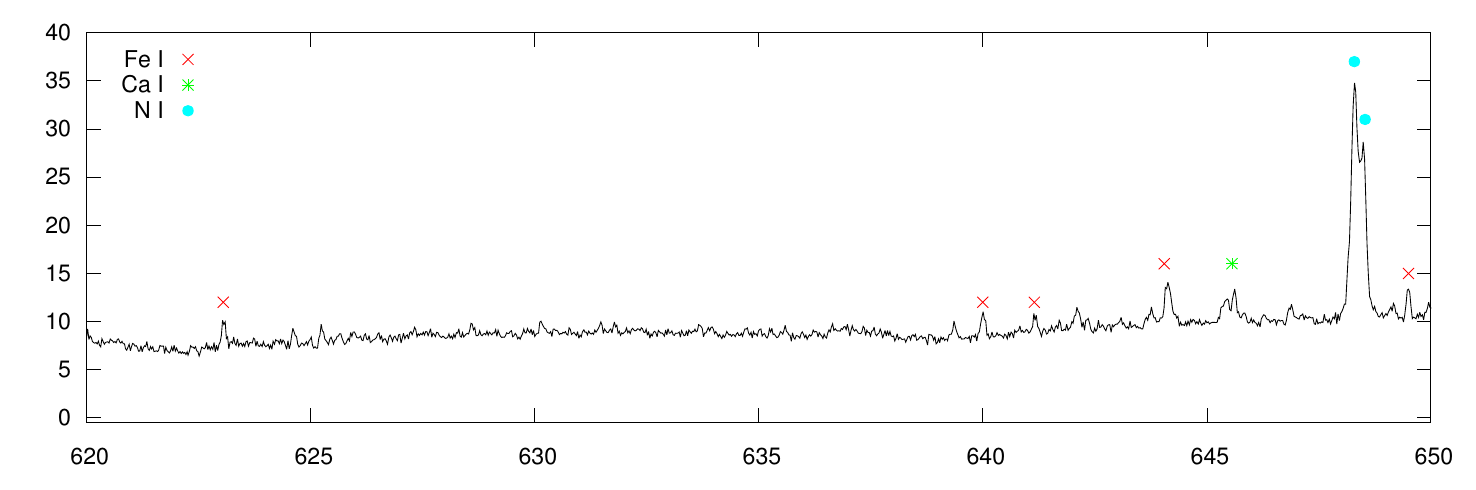}
  \includegraphics[scale=1]{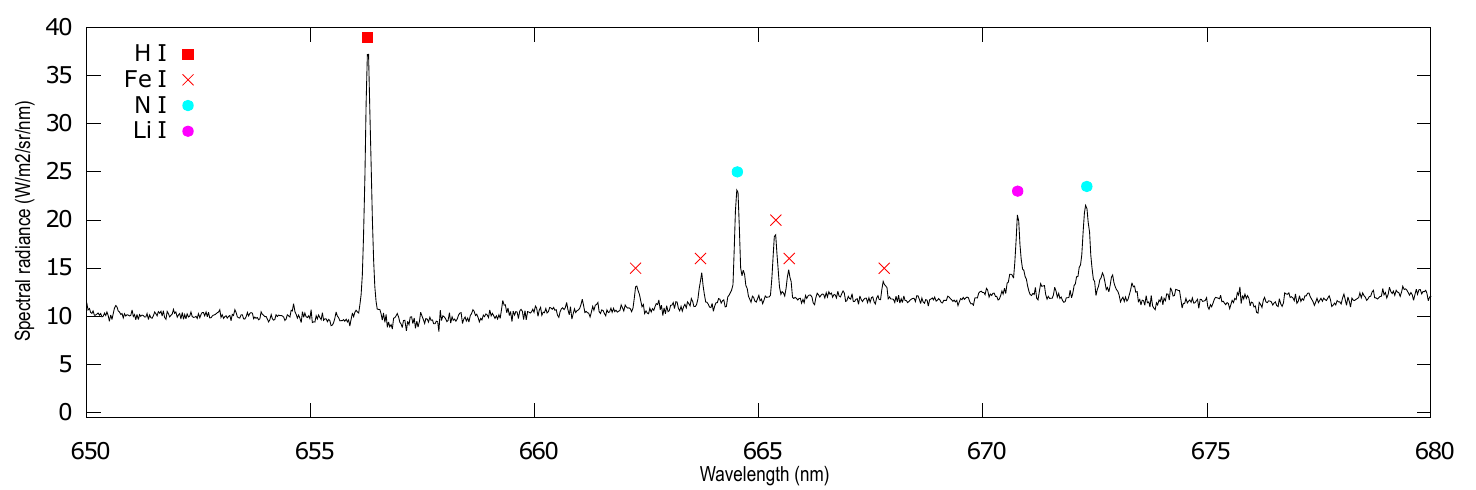}
  \captionsetup{labelformat=empty}
  \caption{Fig A.1 (\emph{Continued})}
\end{figure}
 
\begin{figure}[h!]
\centering  
  \includegraphics[scale=1]{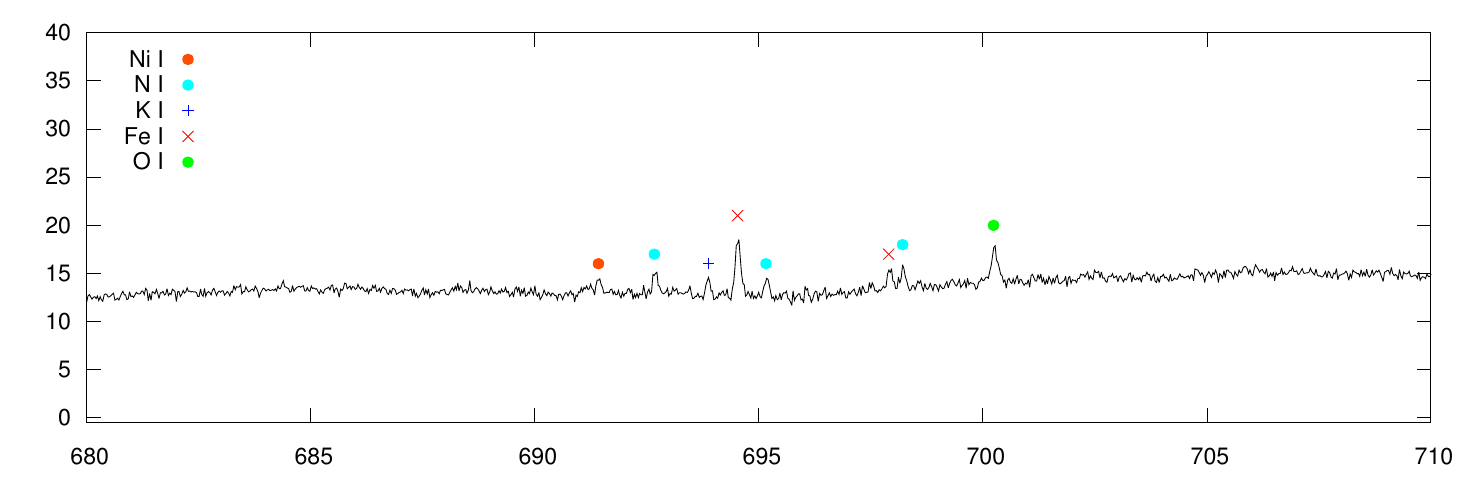}
  \includegraphics[scale=1]{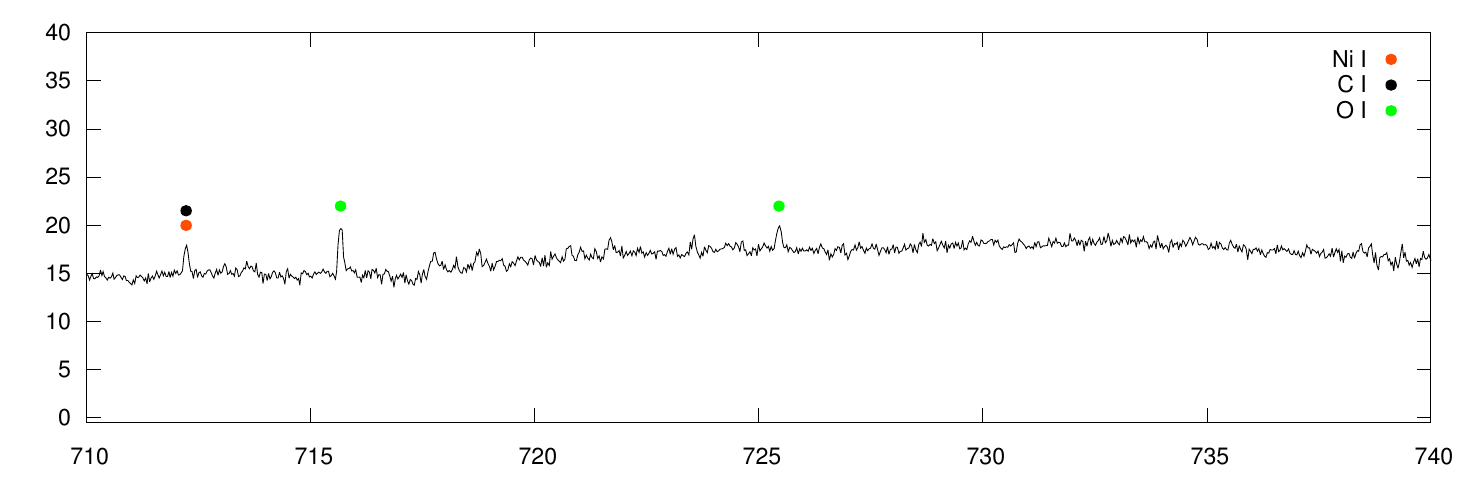}
  \includegraphics[scale=1]{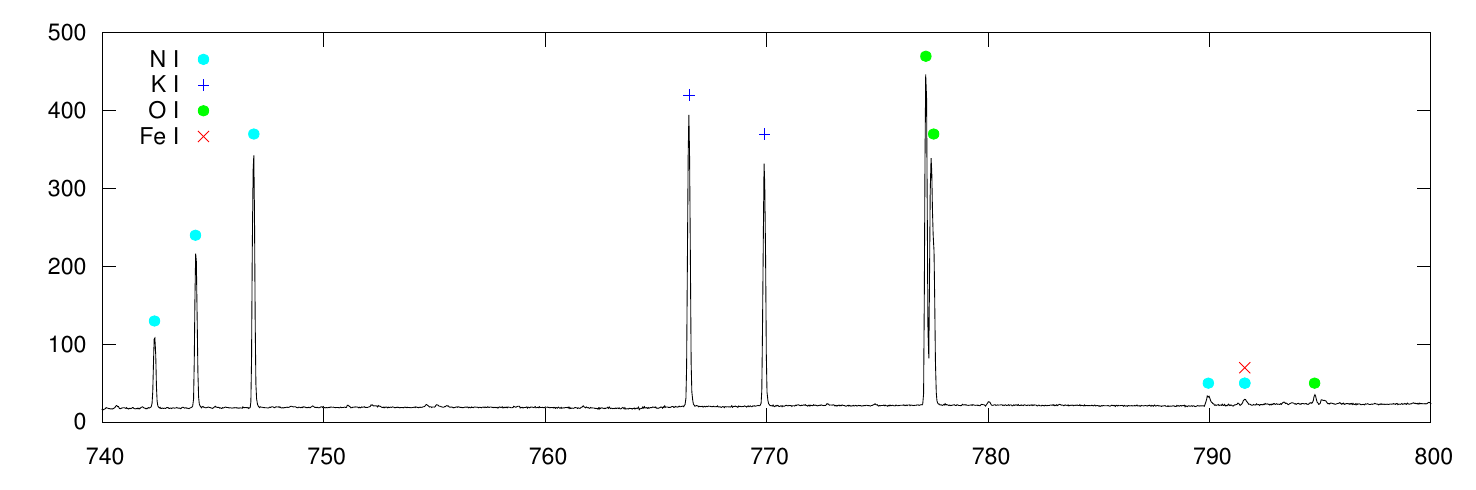}
  \includegraphics[scale=1]{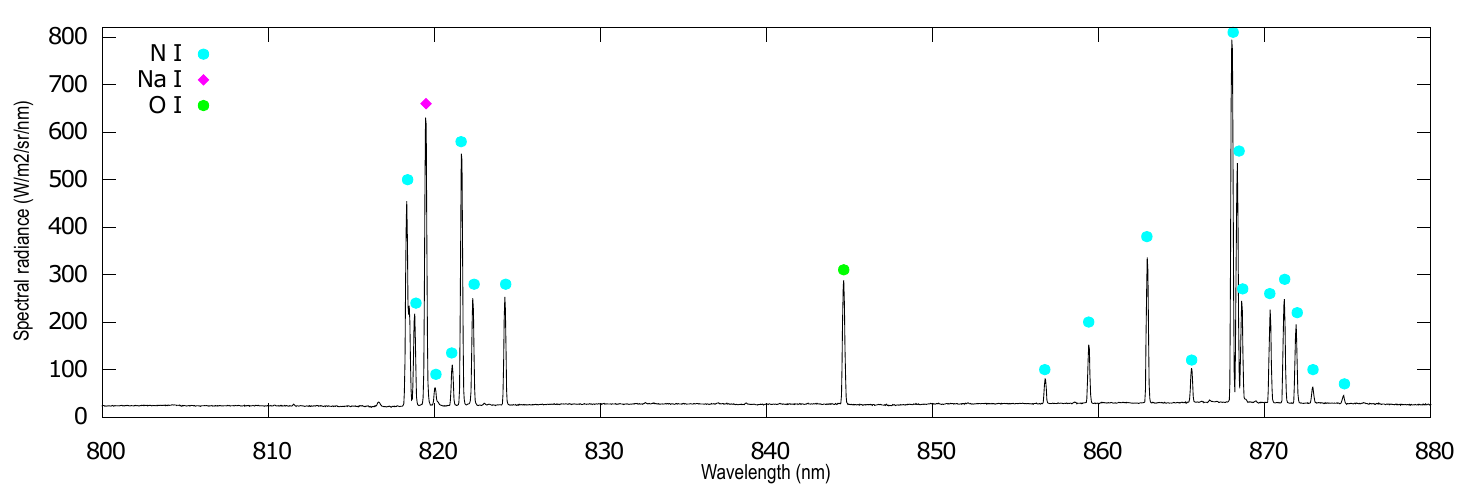}
  \captionsetup{labelformat=empty}
  \caption{Fig A.1 (\emph{Continued})}
\end{figure}

\twocolumn
\onecolumn
\newpage

  \begin{figure}[h!]
\centering
  \includegraphics[scale=0.48]{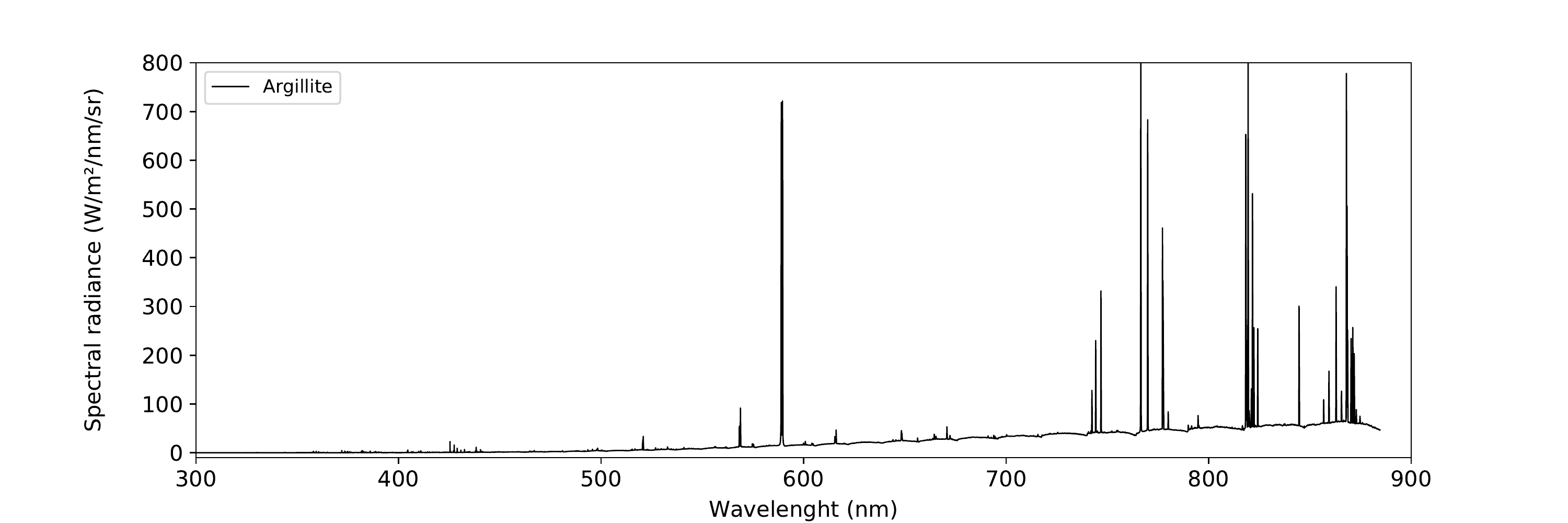}
  \includegraphics[scale=0.48]{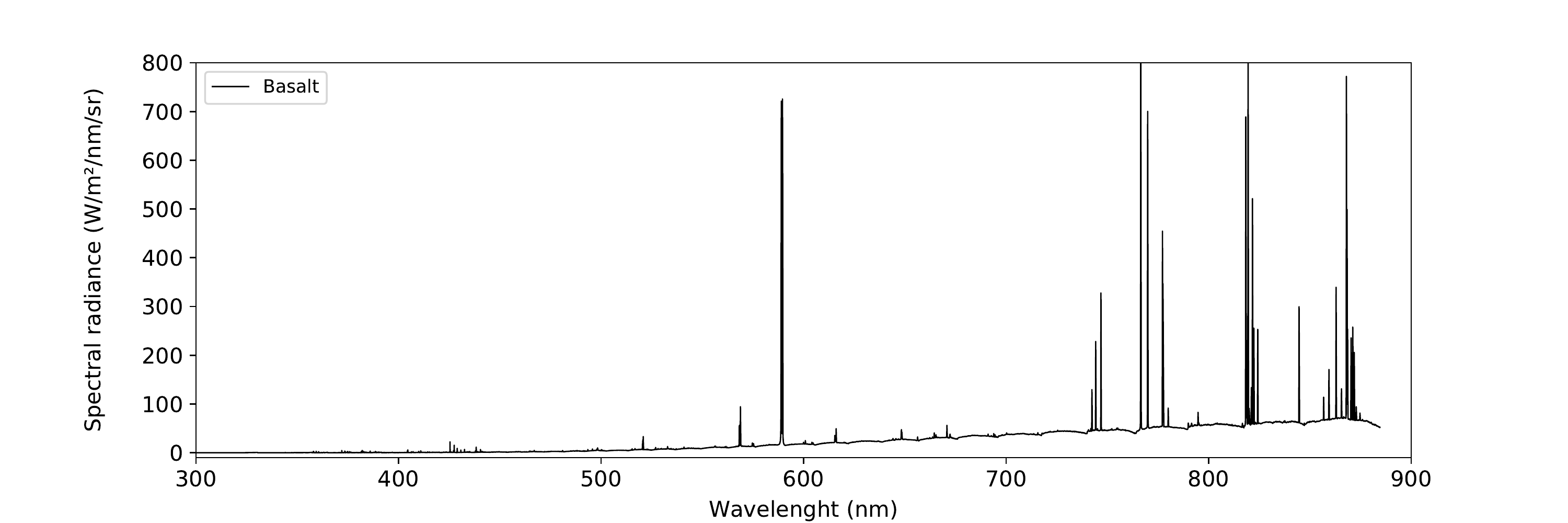}
  \includegraphics[scale=0.48]{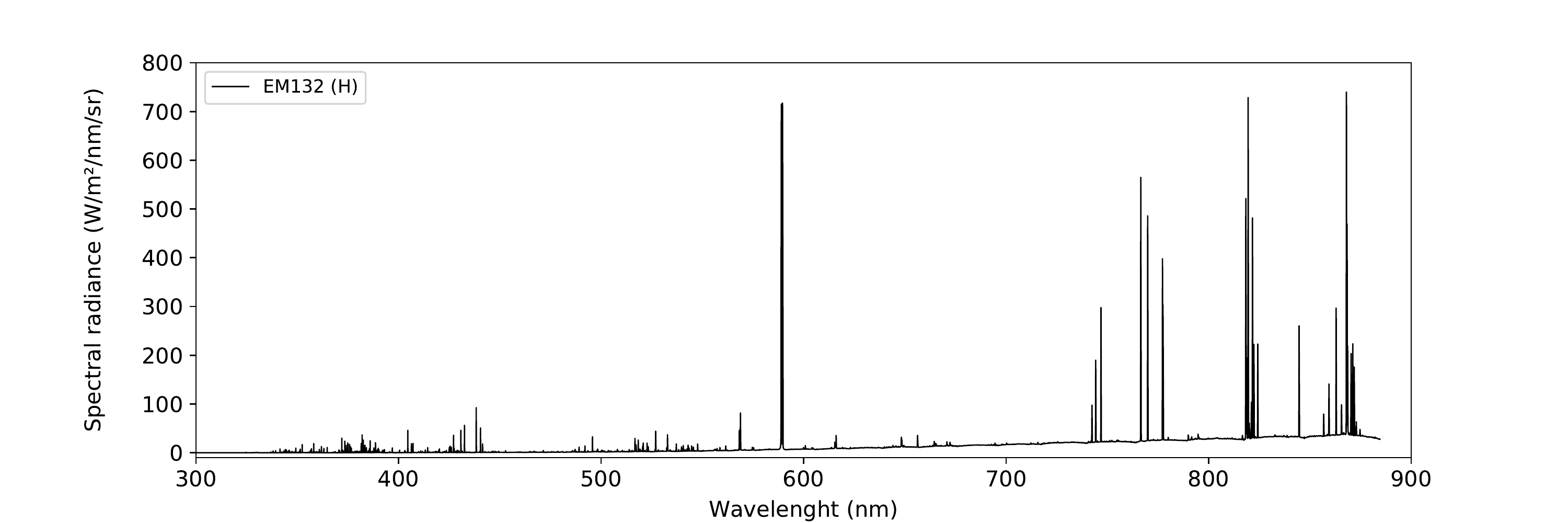}
  \captionsetup{labelformat=empty}
  \caption{Fig. A.2: Spectra collected by \cite{Loehle17} for the H chondrite EM132 and two terrestrial analogues (argillite, basalt).}
  \label{Data_Stefan}
\end{figure}

\twocolumn
\onecolumn
\newpage

\section{Methods of spectral modelling}

Here we present in detail the two methods we used to determine mass ratios from the spectral lines. We considered first a pure emission case in which the emission line intensity is proportional to the abundance as presented in \cite{Nagasawa78} and then we followed an auto-absorbant gas modelling inspired by the model presented in \cite{Borovicka93}. The former considers the gas optically thin whereas the latter considers it optically thick. The first part of this appendix presents the theoretical scheme of radiative transfer and the limit case to model the emission line intensities. The second and third subsections focus on the calculation methods for the two cases (pure emission and auto-absorbant gas model).

\subsection*{B.1: Elements of radiative transfer}

When modelling the ablation process of our meteorite samples, there are three physical processes that need to be taken into account, namely the thermal emission of the sample, the gas emission of the surrounding gas cloud (air enriched in sample compounds by its partial vapourization), and the absorbant effect of the gas with respect to the above emission lines. We detail each of these processes hereafter.\\ 

\noindent
We first assume that the thermal emission of the sample heated by the air friction follows the Planck distribution. The black-body emission at a temperature $T$ is frequency dependent and can be written as  
\begin{equation}
B_\nu (T) = \frac{2h\nu^3}{c^2}\frac{1}{e^{h\nu/ k_B T}-1}
\end{equation}

\noindent
where $\nu$ is the frequency, $c$ the light celerity, $h$ the Planck constant and $k_B$ the Boltzmann constant. The samples were assumed to be grey bodies, implying that their thermal emission $I_\nu^{\mathrm{solid}}$ is a fraction of the Planck function. The emissivity was set at 0.83 for our experimental conditions \citep{Loehle17}.\\

\noindent
As a next step, we modelled the theoretical emission of the hot gas resulting from the sample ablation assuming that\\

\begin{enumerate}[(i)]
\item \label{i} \emph{The gas is in thermal equilibrium.} This classical assumption allows the atomic population derivation and is supported by some theoretical approaches \citep{Boyd00}.
\item \label{ii} \emph{The gas is auto-absorbent without diffusion.} Following the results of \cite{Borovicka93}, we model the gas emission including the auto-absorption.
\item \label{iii} \emph{The line profiles are Voigt profiles.}  A Voigt profile consists of the convolution of a Gaussian profile and a Lorentzian profile. It represents well the two main physical processes involved in the line enlargement, i.e. the Doppler enlargement and the natural enlargement, respectively).
\item \label{iv} \emph{The same profile in emission and absorption.} It implies that the profiles of absorption, spontaneous emission and stimulated emission are the same.
\item \label{v} \emph{The line spread function has a gaussian shape.} The instrumental response consists of the convolution of the modelled intensity with a Gaussian function.
\end{enumerate}

\noindent
In this framework, the gas can be described by the coefficients $\epsilon_\nu$ and $\kappa_\nu$, which are the emission and absorption coefficients of the gas, respectively. Depending on the frequency $\nu$, assuming (\ref{i}) and (\ref{iv}), these coefficients can be written as follows:
\begin{equation}
\epsilon_\nu = \frac{h \nu}{4\pi}\; A_{ul}\; n_u\; \Phi(\nu)
\label{emission}
\end{equation}
\begin{equation}
  \kappa_\nu = \frac{c^2}{8\pi \nu^2} \; A_{ul}\; n_l \frac{g_u}{g_l} \;(1 - e^{- h \nu / k_B T_{gas}})\; \Phi(\nu)
\label{absorption}
\end{equation}

\noindent
where $h$ is the Planck constant, $k_B$ the Boltzmann constant and the following notation:
   \[
      \begin{array}{lp{0.8\linewidth}}
         A_{ul}  & Einstein coefficient of spontaneous emission     \\
         n_u       & upper level population  \\
         n_l       & lower level population  \\
         g_u       & upper level statistic weight\\
         g_l       & lower level statistic weight\\
         \phi(\nu) & line profile            \\
      \end{array}
   \]

\noindent
The emission and absorption coefficients are related to the line profile, assumed to be a Voigt profile (\ref{iii}),

\begin{equation}
  \phi_\nu = \int_{-\infty}^{+\infty} \frac{1}{\pi \Delta\nu_D} e^{-\big( \frac{\nu'-\nu_{ul}}{\Delta\nu_D} \big)^2} * \frac{\delta}{\delta^2 + (\nu-\nu')^2} \mathrm{d}\nu' = H(\Delta\nu_D,\Gamma)
\end{equation}   

\noindent
where $\Delta\nu_D$ is the Doppler width and $\delta$ the lorentzian parameter, defined as follows:

\begin{equation}
\Delta\nu_D = \frac{\nu_{ul}}{c}\sqrt{\frac{2 k_B T_{gas}}{m}} \\
\delta = \frac{1}{4 \pi} \sum_{l<u} A_{ul}=\frac{\Gamma}{4\pi}
\end{equation}

\noindent
where $m$ is the atomic mass of the studied element and $\Gamma$ the damping constant.\\

\noindent
In order to write the radiative transfer equation, we first derived the source function $S_\nu$, ratio of the emission to the absorption coefficient. The source function is, following our assumptions, only temperature dependent because it can be written as a Planck function,

\begin{equation}
  S_\nu= \frac{\epsilon_\nu}{\kappa_{\nu}} = B_{\nu}(T_{\mathrm{gas}})
\end{equation}

      \begin{table}[h!]
      \small
      \caption[]{Main lines for each element with their spontaneous emission coefficient and the upper level degeneracy.}
         \label{lines}
     $$ 
         \begin{array}{p{0.2\linewidth}ccc}
            \hline
            \noalign{\smallskip}
            Element  & \lambda & A_{ul}  &  g_u\\
            \noalign{\smallskip}
            \hline
            \noalign{\smallskip}
Fe I & 381.58 & 1.12\times 10^8 & 7\\
     & 382.78 & 1.05\times 10^8 & 5\\
     & 404.58 & 8.62\times 10^7 & 9\\
     & 438.35 & 5.00\times 10^7 & 11\\
     & 440.48 & 2.75\times 10^7 & 9\\
     & 441.51 & 1.19\times 10^7 & 7\\
\hline
\\   
Ni I & 349.30 & 9.80\times 10^7 & 3\\
     & 547.69 & 9.50\times 10^6 & 3\\
\hline
\\   
Mn I & 403.08 & 1.70\times 10^7 & 8\\
     & 403.31 & 1.65\times 10^7 & 6\\
     & 403.45 & 1.58\times 10^7 & 4\\
\hline
\\
K I  & 766.49 & 3.80\times 10^7 & 4\\
     & 769.90 & 3.75\times 10^7 & 2\\
\hline
\\
Cr I & 425.44 & 3.15\times 10^7 & 9\\
     & 427.48 & 3.07\times 10^7 & 7\\
     & 428.97 & 3.16\times 10^7 & 5\\
     & 520.45 & 5.09\times 10^7 & 3\\
     & 520.60 & 5.14\times 10^7 & 5\\
     & 520.84 & 5.06\times 10^7 & 7\\
\hline
\\
Mg I & 516.73 & 1.13\times 10^7 & 3\\
     & 517.27 & 3.37\times 10^7 & 3\\
     & 518.36 & 5.61\times 10^7 & 3\\
\hline
\\         
Na I & 589.00 & 6.16\times 10^7 & 4\\
     & 589.60 & 6.14\times 10^7 & 2\\
            \noalign{\smallskip}
            \hline
         \end{array}
     $$ 
   \end{table}

\begin{figure} [h!]
   \centering
   \includegraphics[scale=0.46]{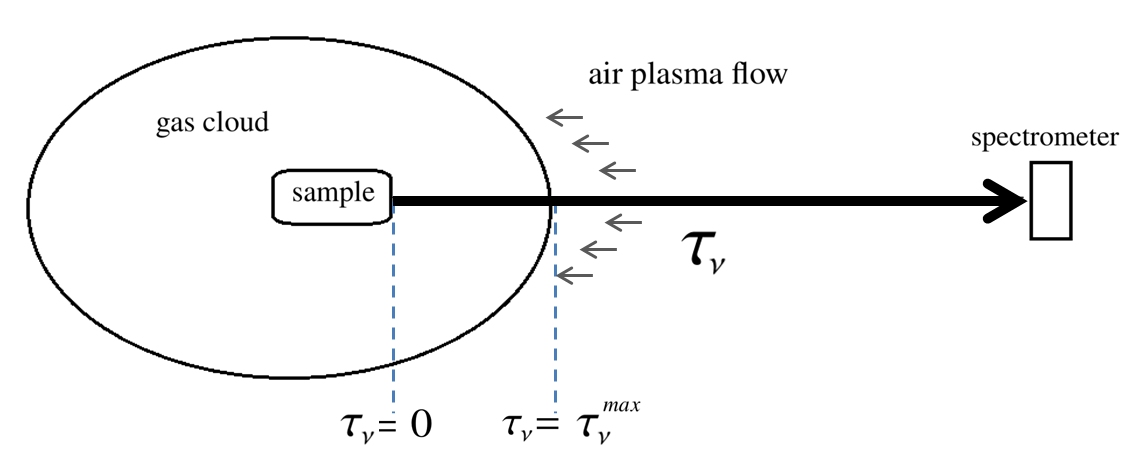}
      \caption{Origin and axis definitions of the optical thickness. $\tau_\nu^{max}$ is defined as the total optical thickness of the cloud. 
              }
         \label{od}
   \end{figure}

\noindent
Let $I_{\nu}$ be the spectral radiance measured by the spectrometer at a given frequency $\nu$. Let $\tau_\nu$ be the optical thickness defined from the sample (see Fig. \ref{od}). The optical thickness is defined by $\tau_\nu = \kappa_\nu \; L$, where $L$ the thickness of the emitting cloud along the line of sight and $\kappa_\nu$ the absorption coefficient. The energy conservation can be written in terms of local variations of the spectral radiance at a given frequency with respect to the optical depth, leading to the following general equation of radiative transfer in a medium without diffusion:
\begin{equation}
  \frac{\mathrm{d}I_{\nu}}{\mathrm{d}\tau_\nu}= S_{\nu}-I_{\nu}
\end{equation}

\noindent
The general solution can be written as
\begin{equation}
  I_\nu(\tau_\nu) = I_\nu(\tau_\nu = 0) \; e^{-\tau_\nu} + \int_0^{\tau_\nu} S_\nu(\tau_\nu') \; e^{\tau_\nu - \tau_\nu'} \; \mathrm{d}\tau_\nu'
\end{equation}

\noindent 
Following previous assumptions, $I_\nu(\tau_\nu = 0)=I_\nu^{\mathrm{sample}}$. Moreover, because the source function is temperature dependent only, it is uniform for the optical thickness and: 
\begin{equation}
  I_\nu(\tau_\nu) = I_\nu^{\mathrm{solid}}\; e^{-\tau_\nu} + B_\nu(T_{\mathrm{gas}}) (1-e^{-\tau_\nu})
\label{transfer}
\end{equation}

\noindent
As expected, the spectral radiance measured in the spectrometer is the addition of the thermal continuum of the sample and the spectral lines of the hot surrounding gas cloud. The intensity collected in the spectrometer is simply $I_\nu(\tau_\nu^{max})$, defined in Fig. \ref{od}, where $\tau_\nu^{max} = \kappa_\nu L$, $L$ is the cloud radius. The optical thickness was set at $L=30.0 \; cm$, a reasonable distance considering the dimensions of the facilities. 

\noindent
We finally can write the spectral line intensity $I^{\mathrm{thin}}_\nu(\tau_\nu)$ and $I^{\mathrm{thick}}_\nu(\tau_\nu)$ for a gas optically thin (section 3.2.1) or thick (section 3.2.2), respectively, as

\begin{equation}
  I^{\mathrm{thin}}_\nu(\tau_\nu) = \lim_{\tau_\nu \to 0} I_\nu(\tau_\nu) =\frac{h \nu}{4\pi}\; A_{ul}\; n_u\; L \; \Phi(\nu)
\label{Thin}
\end{equation} 

\begin{equation}
  I^{\mathrm{thick}}_\nu(\tau_\nu) = I_\nu^{\mathrm{sample}}\; e^{-\tau_\nu} + B_\nu(T_{\mathrm{gas}}) (1-e^{-\tau_\nu})
 \label{Thick}
\end{equation}   

\subsection*{B.2: Optically thin case}

In such a case, the integrated intensity along the considered spectral line is proportional to the upper level population following equation \ref{Thin} and therefore to the atomic population in the gas using the Boltzmann distribution 

\begin{equation}
n_u = n \; \frac{g_u}{U} e^{-\Phi_u/ k_B T}
\end{equation}

\noindent
This implies a proportionality between intensity ratios and mass ratios. We can therefore compare intensity ratios between various spectra using the same spectral lines. In the present study, we derived two intensity ratios (Mg/Fe, Na/Fe) that we weighted by the atomic masses. For example, in the case of Mg/Fe, the procedure is as follows:

\begin{equation}
r_{Mg/Fe} = \frac{I^{Mg} m^{Mg}}{I^{Fe} m^{Fe}} 
\label{massratio}
\end{equation}
   
\subsection*{B.3: Optically thick case}

\begin{figure} [h!]
   \centering
   \includegraphics[scale=0.7]{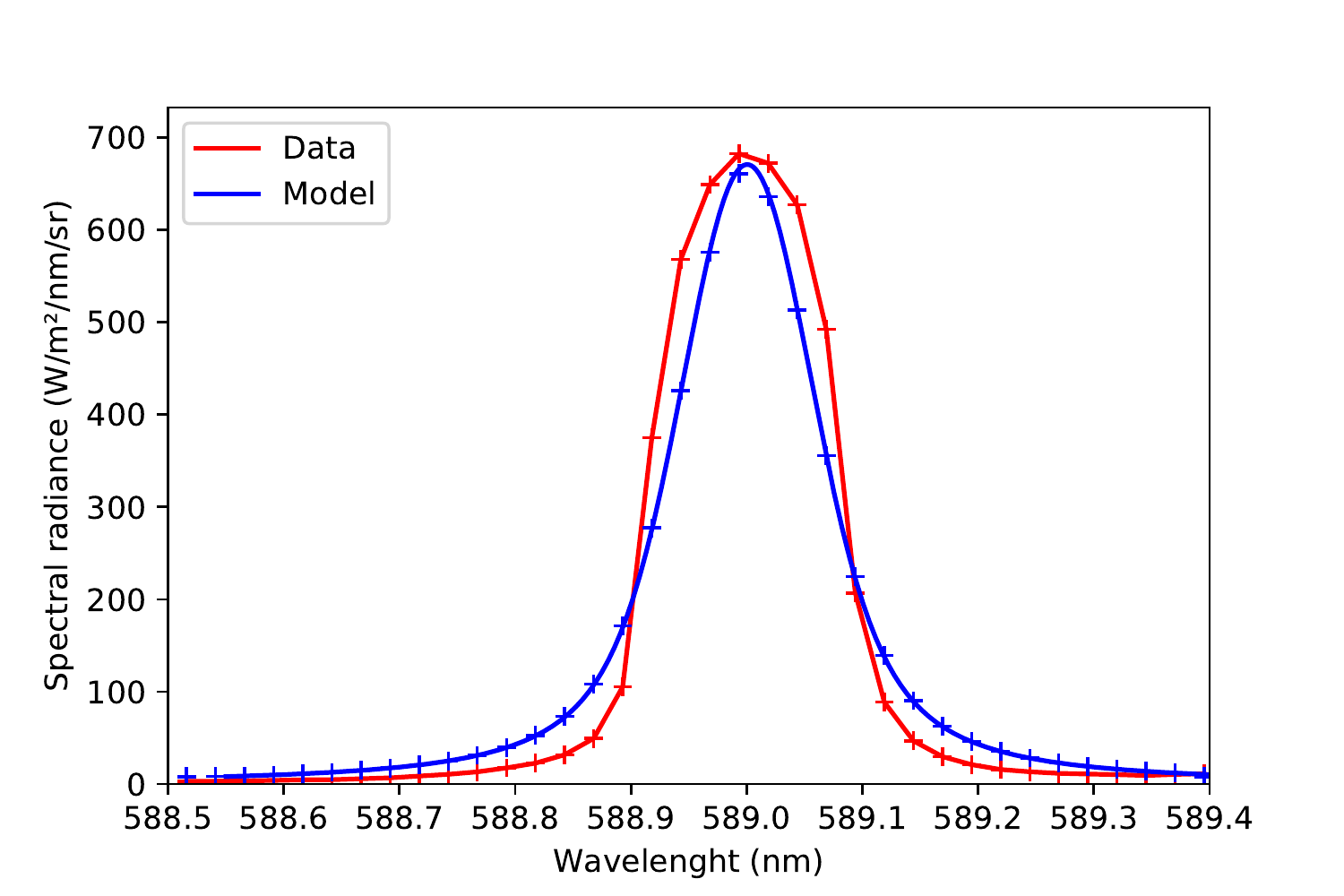}
      \caption{Sodium line (589.0 $nm$) of the sample EM132 plotted with its modelled line (outputs: T=3200 $K$ and n=9.8$\times$10$^{14}$).
              }
         \label{fit}
   \end{figure}

Considering an optically thick case, the intensity is described by equation \ref{Thick} and has to be integrated numerically. We remind here that the area under an emission line depends only on two free parameters, namely the temperature and atomic abundance in the lower level (see first section). To derive both parameters, we minimized the integrated intensity difference between the experimental data and the model points using the least-squares fitting method (Fig. \ref{fit}). We applied this fitting technique to several iron emission lines and determined the gas temperature in this way. We then kept this temperature fixed and ran the model on the lines of interest (Na, Mg, and Fe) to determine the atomic abundance of these species in the lower level.\\ 

\noindent
In order to perform the least-squares method with the data, we convolved the computed intensity $I_\nu(\tau_\nu^{max})$ with the instrumental response (a Gaussian following (\ref{v})). The synthetic spectral line intensity ($\mathcal{I}_\nu^{\mathrm{synth}}$) is therefore expressed as:\\ 

\begin{equation}
  \mathcal{I}_\nu^{\mathrm{synth}} = \int_{-\infty}^{+\infty} I_\nu(\tau_\nu^{max}) * e^{-\big( \frac{\nu'-\nu_{ul}}{2 \sigma} \big)^2}  \mathrm{d}\nu'
\end{equation} 
\noindent
where  $\sigma$ is the full width at half-maximum and corresponds to the spectral distance between two pixels.\\

\noindent
As a next step, we constrained for each element (Na, Mg, and Fe) the total atomic population present in the gas using the Boltzmann distribution that allows us to link the lower level population $n_l$ to the atomic population $n$ via the following relation:

\begin{equation}
n_l = n \; \frac{g_l}{U} e^{-\Phi_l/ k_B T}
\end{equation}

\noindent
Finally, we derived the mass ratios (Mg/Fe, Na/Fe) using the previously estimated atomic populations that we weighted by the atomic mass. For example, in the case of Mg/Fe, the procedure is as follows:

\begin{equation}
r_{Mg/Fe} = \frac{n^{Mg} m^{Mg}}{n^{Fe} m^{Fe}}
\label{massratio}
\end{equation}

\newpage
\section{Additional tables}

\begin{table}[h!]
\caption{Mg/Fe}
\centering
\begin{tabular}{lccc}
\hline
Sample & Mean bulk & Intensity ratio & Spectral modelling \\
\hline
Agen       & 0.51 & 0.078 $\pm$ 0.01 & 0.004 $\pm$ 0.002 \\
Argilitte  & 0.92 & -                & -     \\
Basalt     & 0.76 & -                & -     \\
EM132      & 0.51 & 0.11 $\pm$ 0.01  & 0.14 $\pm$ 0.01  \\
Granes     & 0.69 & 0.08 $\pm$ 0.01  & 0.045 $\pm$ 0.008 \\
Juvinas    & 0.29 & 0.04 $\pm$ 0.01  & 0.028 $\pm$ 0.008     \\
Murchison  & 0.58 & 0.07 $\pm$ 0.01  & 0.051 $\pm$ 0.008 \\
St-Séverin & 0.80 & 0.07 $\pm$ 0.01  & 0.051 $\pm$ 0.008 \\
\hline
\end{tabular}
\end{table}

\begin{table}[h!]
\caption{Na/Fe}
\centering
\begin{tabular}{lccc}
\hline
Sample & Mean bulk & Intensity ratio & Spectral modelling \\
\hline
Agen       & 0.023 & 1.3 $\pm$ 0.2 & 0.01 $\pm$ 0.09 \\
Argilitte  & 0.036 & -              & -      \\
Basalt     & 0.33 & 24.3  $\pm$ 0.6 & 5.1 $\pm$ 0.06   \\
EM132      & 0.023 & 2.9 $\pm$ 0.2 & 0.16 $\pm$ 0.07  \\
Granes     & 0.033 & 8.1 $\pm$ 0.3 & 0.51 $\pm$ 0.08  \\
Juvinas    & 0.021 & 3.8 $\pm$ 0.1 & 0.20 $\pm$ 0.07  \\
Murchison  & 0.020 & 3.6 $\pm$ 0.1 & 0.16 $\pm$ 0.07  \\
St-Séverin & 0.038 & 4.1 $\pm$ 0.2 & 0.26 $\pm$ 0.07  \\
\hline
\end{tabular}
\end{table}

\end{appendix}

\end{document}